\documentclass[aps,pra,onecolumn,12pt,notitlepage,floatfix,amsmath,amssymb]{revtex4-2}
\usepackage{graphicx}
\usepackage[T1]{fontenc}
\usepackage{lmodern}
\usepackage{url}
\usepackage{xcolor}
\usepackage[colorlinks=true,linkcolor=blue!60!black,citecolor=blue!60!black,urlcolor=blue!60!black]{hyperref}
\usepackage{orcidlink}
\hypersetup{pdftitle={Potential Flow, Stress, and Straightness in Flow Matching: Euler Error on Gaussian-Mixture Targets},pdfauthor={Zijian Liu, Julian Olszewski, Jie Xu},pdfsubject={Flow matching and Euler discretization}}
\providecommand{\tightlist}{\setlength{\itemsep}{0pt}\setlength{\parskip}{0pt}}
\makeatletter
\renewcommand{\p@subsection}{}
\makeatother

\renewcommand{\thetable}{\arabic{table}}

\begin{document}
\title{Potential Flow, Stress, and Straightness in Flow Matching: Euler Error on Gaussian-Mixture Targets}
\author{Zijian Liu\,\orcidlink{0009-0000-6014-1268}}
\thanks{Z. Liu and J. Olszewski contributed equally to this work.}
\affiliation{Department of Mechanical and Industrial Engineering, University of Illinois Chicago}
\author{Julian Olszewski\,\orcidlink{0009-0006-2346-8097}}
\thanks{Z. Liu and J. Olszewski contributed equally to this work.}
\affiliation{Department of Mechanical and Industrial Engineering, University of Illinois Chicago}
\author{Jie Xu\,\orcidlink{0000-0001-5765-7431}}
\email[Corresponding author: ]{jiexu@uic.edu}
\affiliation{Department of Mechanical and Industrial Engineering, University of Illinois Chicago}
\date{12 November 2026}
\begin{abstract}
In linear flow matching, coupling straightness measures conditional displacement variance. The conditional covariance gives a kinetic stress in the Jeans balance. The virial identity relates the mean trace of the covariance to a radial moment of acceleration. Euler error also depends on how the flow propagates acceleration. On constructed Gaussian-mixture targets, closed-form couplings with matched straightness yield different median Euler errors. Learned fields from fresh minibatch assignments also yield different Euler errors from their partners at matched estimated straightness. The leading Euler coefficient accounts for more of the error variation than straightness in these experiments. We relate the additive-noise characterization of isotropic Gaussian sources to potential flow. A gradient constraint reduces velocity error on the independent isotropic Gaussian example and increases it on two examples with rotational exact fields.
\end{abstract}

\maketitle

\section{Introduction and governing identities}\label{sec:1}

Flow matching learns a velocity field and integrates its ordinary differential equation \cite{r1,r2}. The pairing of source and target samples influences the field and the error made when integrating it with a fixed number of Euler steps. Rectified flow and minibatch transport use this choice to encourage straighter trajectories \cite{r2,r3,r4,r5}.

Let \((x_0,x_1)\) follow a coupling \(\pi\) of a source \(p_0\) and a target \(p_1\) on \(\mathbb R^d\). Set \(x_t=(1-t)x_0+tx_1\), \(u=x_1-x_0\), and write \(\rho_t\) for the density of \(x_t\). The regression field is \(v(x,t)=E[u\mid x_t=x]\). Define \(\Sigma=\mathrm{Cov}[u\mid x_t=x]\), \(J_{ij}=\partial_jv_i\), and \(a=\partial_tv+Jv\). Write \(X_t\) for a sample of \(\rho_t\) and \(E_t\) for its expectation. We call \(S=\int_0^1 E_t\,\mathrm{tr}\,\Sigma\,dt\) \emph{coupling straightness}. It equals Liu et al.'s crossing loss \(V(X)\) \cite[Eq.~12]{r2} and Lee et al.'s degree of intersection \(I(q)\) \cite[Eq.~7]{r5}. The ODE chord-deviation statistic uses the endpoints of the ODE's own trajectories \cite[Eq.~3]{r2}\cite[Sec.~3]{r4}. Pham et al.~\cite[Theorem 5.1]{pham} bound the crossing loss of the reflow ODE's endpoint coupling by a multiple of its trajectory straightness defect when the defect is finite and the velocity is uniformly Lipschitz in space.

Tao and Choi found that their straightest learned pinwheel trajectories did not have the smallest same-field Euler error \cite[Fig.~5]{r6}. Claustre et al.~\cite[Proposition 1]{claustre} show that for independent centered Gaussians with commuting covariances, the interpolation schedule can vary regression variance at fixed marginal path and drift. Yin \cite[Proposition 1]{r11} studies propagation of the local defect, and Obzherin and Khrulkov \cite{r17,r18} analyze its effect on the output law.

Bo\"it\'e et al.~\cite[Sec.~4.4]{boite} compare mean Euler endpoint errors relative to the corresponding exact flow as batch size and step count vary for semidiscrete targets. Chen and Levine \cite[Proposition 4.13 and Sec.~4.2]{chen} bound the integrated regression variance of unregularized minibatch optimal transport by the excess quadratic transport cost. Their one-step distributional error uses squared Wasserstein distance.

We fix the linear interpolation schedule and compare median Euler endpoint errors at matched estimates of \(S\) for couplings with a shared smooth mixture target. We use the Jeans balance to connect \(S\) to a radial acceleration moment and express the weak Euler coefficient in terms of conditional covariance. For independent interpolation, we use the two-point condition from Gribonval and Nikolova \cite[Lemma 3(b)]{gribonval} to characterize sources that yield potential flow for every two-point target. We test gradient constraints on examples with potential and rotational exact fields.

\subsection{Jeans balance and virial identity}

Each sample pair moves ballistically along its linear interpolant. At a given position, different pairs can carry different velocities. The Jeans balance relates the material acceleration of the mean flow to the divergence of the resulting kinetic stress.

In the compact-domain setting of Tsimpos and Marzouk \cite[Lemma 3 and Sec.~2.3]{r20}, the momentum balance has the conditional mean of the path acceleration on its right-hand side. For our linear interpolant \(x_t\), each path has constant velocity \(u\), so that term is zero and their Reynolds tensor is \(\Sigma\). Where the marginal fields are smooth and \(\rho>0\), continuity and the Jeans balance give
\begin{equation}
\label{eq:jeans}
\partial_t\rho+\nabla\!\cdot(\rho v)=0,\qquad \rho a=-\nabla\!\cdot(\rho\Sigma).
\end{equation}
The momentum equation is also given by Khan \cite[Appendix A.5, Eq.~35]{r15}. Here \(\rho\Sigma\) is the kinetic stress tensor, as in the Jeans equations \cite[Sec.~4.8]{r14}.

When the Jeans balance holds on \(\mathbb R^d\) in distributions, for a smooth test field \(\xi\) with \(E_t[|\Sigma||\xi|]<\infty\) and both integrands integrable, testing the balance gives
\begin{equation}
\label{eq:tested-jeans}
E_t[\xi\cdot a]=E_t[\Sigma:\nabla\xi].
\end{equation}
For \(\xi=x\), finite second moments and \(E_t|X_t\cdot a|<\infty\) suffice for
\begin{equation}
\label{eq:virial}
E_t\,\mathrm{tr}\,\Sigma=E|u|^2-E_t|v|^2=E_t[X_t\cdot a].
\end{equation}
In particular \(E_t\,\mathrm{tr}\,\Sigma\le(E_t|X_t|^2)^{1/2}(E_t|a|^2)^{1/2}\). Tsimpos and Marzouk use the same radial test to relate the stress trace to the radial acceleration of sample paths when the marginal flow is straight \cite[Theorem 6]{r20}.

\subsection{Potential flow and Gaussian covariance}

Fix \(0<t<1\). For independent \(x_0\sim N(\mu_0,S_0)\) with positive definite covariance and any target of finite second moment,
\begin{equation}
\label{eq:gaussian-score}
v=\frac{x-\mu_0}{t}+\frac{1-t}{t}S_0\nabla\log\rho_t.
\end{equation}
The Jacobian is symmetric where \(S_0\) commutes with \(\nabla^2\log\rho_t\). For \(S_0=\sigma^2 I\) this holds everywhere and \(v=\nabla\Phi\), with \(\Phi=|x-\mu_0|^2/(2t)+\frac{(1-t)\sigma^2}{t}\log\rho_t\) \cite[Eq.~3.9]{gao}\cite[Example 5.1]{r16}\cite[Proposition 3.1]{r22}. The covariance-dependent form follows from multivariate Tweedie regression \cite[Eq.~3]{dytso}. Conversely, let \(d\ge2\) and let the independent source have an everywhere positive \(C^2\) density. If the field is curl-free at that time for every two-point target, then the source is an isotropic Gaussian. The related denoising characterization treats posterior means that are proximity operators \cite[Lemmas 3--4 and Proposition 4]{gribonval}. For differentiable schedules \(x_t=\alpha_tx_1+\beta_tx_0\) with an independent \(x_0\sim N(0,\sigma^2I)\), at times with \(\alpha_t\beta_t\ne0\) the velocity follows from the Gaussian schedule formula
\begin{equation}
\label{eq:gaussian-schedule}
v=\frac{\dot\alpha}{\alpha}x+\left(\frac{\dot\alpha\beta^2}{\alpha}-\dot\beta\beta\right)\sigma^2\nabla\log\rho_t.
\end{equation}
This is the schedule formula of Albergo et al.~\cite[Eqs.~4.16 and 4.19]{albergo}, with their data and noise coefficients set to \(\alpha_t\) and \(\sigma\beta_t\), respectively.

For \(0<t<1\), use the posterior covariance identities of Gao et al.~\cite[Eq.~4.3]{gao}, Wan et al.~\cite[Appendix C.1, Proposition C.2]{wan}, and Xing et al.~\cite[Corollary 3, Eq.~21]{r7} in coordinates translated by \(\mu_0\) and scaled by \(\sigma\). The velocity Jacobian is unchanged by this transformation, and \(u=(x_1-x)/(1-t)\) at fixed \(x_t=x\). Restoring displacement units gives
\begin{equation}
\label{eq:conditional-covariance}
\Sigma=\frac{\sigma^2}{t}\big(I+(1-t)J\big).
\end{equation}
Substitution into the Jeans balance~\eqref{eq:jeans} gives
\begin{equation}
\label{eq:gaussian-jeans}
\rho a=-\frac{\sigma^2}{t}\big[\nabla\rho+(1-t)\nabla\!\cdot(\rho\nabla v)\big].
\end{equation}
Bansal et al.~\cite[Appendix D.1, Eq.~23]{bansal} give its expansion in derivatives of the log density for the standard Gaussian source.

\subsection{Euler error and coupling straightness}

Let \(Y_t\) follow the exact marginal ODE from \(Y_0\sim p_0\), and let \(\hat Y_N\) be the Euler endpoint from the same start with step \(h=1/N\). With smoothness through \(t=1\) and bounded derivatives along the exact and numerical paths, the endpoint error satisfies
\begin{equation}
\label{eq:euler-trajectory}
\hat Y_N-Y_1=-\frac h2 g(1)+O(h^2),\qquad \dot g=J(Y_t,t)g+a(Y_t,t),\quad g(0)=0.
\end{equation}
This is the classical principal global error term for explicit Euler \cite[Sec.~II.8]{r24}. For a smooth observable \(\psi\) and \(u_t=\psi\circ\Phi_{t\to1}\), assume integrable Taylor remainders and that \(\xi=\nabla u_t\) satisfies the integrability conditions for Eq.~\eqref{eq:tested-jeans}, with \(E_t[|\nabla u_t\cdot a|+|\Sigma:\nabla^2u_t|]\) integrable in time. Then
\begin{equation}
\label{eq:euler-weak}
E\psi(\hat Y_N)-E\psi(Y_1)
=-\frac h2\int_0^1 E_t[\Sigma:\nabla^2u_t]\,dt+O(h^2).
\end{equation}
Coupling straightness averages \(\mathrm{tr}\,\Sigma=\Sigma:I\) over space and time. In Eq.~\eqref{eq:euler-weak}, the corresponding contraction is with the Hessian \(\nabla^2u_t\) of the transported observable.
Obzherin and Khrulkov give the density-level Euler defect and its propagation \cite[Theorems 1 and 3]{r17}. Their local mixture comparison gives the covariance deficit of the barycentric step \cite[Appendix D.3, Theorem 36]{r18}.

For any function \(v_\theta(x,t)\), the standard least-squares decomposition gives
\begin{equation}
\label{eq:loss-decomposition}
\int_0^1E|u-v_\theta(x_t,t)|^2\,dt=S+\int_0^1E|v_\theta-v|^2\,dt.
\end{equation}
The residual \(u-v\) is orthogonal in \(L^2\) to every function of \((x_t,t)\) and \(E|u-v|^2=E\,\mathrm{tr}\,\Sigma\). For this regression the parameter-independent loss offset is \(S\), as in the conditional and marginal loss relation of Lipman et al.~\cite[Theorem 2]{r1}. Lee et al.~\cite[Proposition 1]{r5} use the corresponding upper bound on \(I(q)\), and Pooladian et al.~\cite[Theorem 4.2]{r4} study the optimal conditional loss under minibatch couplings.

For quadratic population optimal transport, zero conditional covariance gives zero acceleration by the Jeans balance~\eqref{eq:jeans}. The potential property in that setting follows separately from the optimal map \cite[Theorems 4 and 5]{r6}. Fresh finite minibatches define the expected assignment coupling \cite[Definition 3.1]{boite}.

\section{Radial acceleration, the potential converse and weak Euler error}\label{sec:2}

To obtain the virial identity, test the Jeans balance~\eqref{eq:jeans} against \(\xi\) with a smooth radial cutoff \(\chi_R(x)=\chi(|x|/R)\), equal to one on \(|x|\le R\) and zero beyond \(2R\). Integration by parts gives \(E_t[\chi_R\xi\cdot a]=E_t[\chi_R\Sigma:\nabla\xi]+E_t[\Sigma:(\xi\otimes\nabla\chi_R)]\). The final term tends to zero under the stated integrability condition. For \(\xi=x\), \(|x||\nabla\chi_R|\) is bounded independently of \(R\) on the annulus \(R\le |x|\le 2R\). Since \(E_t\,\mathrm{tr}\,\Sigma\le E|u|^2\), the annular term tends to zero with finite second moments. Dominated convergence gives \(E_t[X_t\cdot a]=E_t\,\mathrm{tr}\,\Sigma\). The other equality is the law of total variance.

At a fixed interior time, \(v(x,t)=(E[x_1\mid x_t=x]-x)/(1-t)\). Thus symmetry of the FM Jacobian is equivalent to symmetry of the posterior-mean Jacobian in the additive model \(x_t=tx_1+(1-t)x_0\). For an equally weighted two-point target, differentiating the posterior weights reduces symmetry to collinearity of the source score difference with the separation of the two translated source arguments. The same collinearity condition appears in Gribonval and Nikolova \cite[Lemma 3(b), Eq.~21]{gribonval}, where it follows from the proximity-operator assumption. In the limit of coincident arguments the Hessian of the log source density is a scalar multiple of the identity. In \(d\ge2\), its vanishing mixed partials make that scalar constant, and integrability forces it to be negative. The Gaussian score formula~\eqref{eq:gaussian-score} gives sufficiency.

Along \(Y_t\), the adjoint gradient \(\nabla u_t(Y_t)\) solves \(d(\nabla u_t(Y_t))/dt=-J^T\nabla u_t(Y_t)\). Pairing it with \(g\) and integrating gives \(E[\nabla\psi(Y_1)\cdot g(1)]=\int_0^1 E_t[\nabla u_t\cdot a]\,dt\). The tested balance~\eqref{eq:tested-jeans} replaces the integrand by \(E_t[\Sigma:\nabla^2u_t]\).

\section{Numerical checks of the identities}\label{sec:3}

We check the Jeans balance against closed-form fields, point-mass targets, direct quadrature and a nonlinear three-stream coupling. The residuals and source controls distinguish numerical accuracy from the effects of changing the source law (Supplementary Tables~\ref{tab:1}, \ref{tab:2} and \ref{tab:3}). Sec.~\ref{supp:checks} gives the numerical settings.

The independent isotropic Gaussian cases give a curl-free field to numerical accuracy. The anisotropic and non-Gaussian controls can have nonzero curl. The commuting Gaussian example shows that an anisotropic source can still give a curl-free field for a particular target.

The principal Euler term captures the observed weak error in the smooth test cases. The conditional-covariance integrand can have high variance near separatrices (Sec.~\ref{supp:checks}).

\section{Straightness and Euler error}\label{sec:4}
We start with matched closed-form couplings, then use minibatch assignment frequencies to construct closed-form cell plans. Finally, we train velocity fields directly on fresh minibatch assignments and compare them with matched partners. At each step count, we measure the median Euclidean distance from the Euler endpoint to a reference endpoint for the same field.

\subsection{Closed-form couplings with a common target}\label{sec:closed}

For each of 11 target seeds (21 to 31), we draw a five-component Gaussian mixture \(T_0\) in the plane. We integrate the independent flow from \(N(0,I)\) to \(T_0\) by 300-step RK4 to obtain the map \(\phi\), and differentiate the numerical map. A lattice with 759 centers \(\mu_i\), weights \(c_i\), and width \(s=0.4\) approximates the source Gaussian. We define the target by the image of each cell under its linearized map,
\[p_1=\sum_ic_i\,N\big(\phi(\mu_i),\,s^2D\phi_iD\phi_i^T\big),\]
IND and LAM use \(N(0,I)\) as the source, DIAG uses the finite Gaussian lattice, and MIXRF mixes these source laws. The lattice and standard Gaussian agree at the probed locations to the numerical accuracy reported in Sec.~\ref{supp:closed}. All families share the constructed target before numerical pruning. We use 14 couplings per target:

\begin{itemize}
\tightlist
\item
  IND: the independent coupling of \(p_0\) and \(p_1\).
\item
  LAM(\(\lambda\)), \(\lambda=0.5,0.8,1\): Gaussian pairs correlated within each target component, \(\mathrm{Cov}(x_0,x_1\mid j)=\lambda B_j^{1/2}\) for target component \(j\) with covariance \(B_j\).
\item
  DIAG(\(\kappa\)), \(\kappa=0,0.5,0.8,1\): source cell \(i\) paired with target component \(i\), \(\mathrm{Cov}(x_0,x_1\mid i)=\kappa\,s^2D\phi_i^T\). DIAG(1) is the affine cell map (the reflow-like pairing). On one target the optimal-transport plan between the cell sets, from a linear program with a dual certificate over all \(576081\) edges, has \(S\) within 0.4 percent of it, a comparison of this scalar on one target. DIAG(0) is independent inside each pair.
\item
  MIXRF(\(\theta\)): with probability \(\theta\) the pair follows DIAG(1), otherwise IND. We tune \(\theta\) to match the quadrature estimate of \(S\) for each of the six LAM and DIAG(\(\kappa<1\)) couplings. A mixture of couplings has \(\rho\), \(\rho v\) and \(\rho M\) (\(M=\Sigma+vv^T\)) as convex combinations of those of its parts, so \(S(\theta)\) is evaluated from the stored per-point values of IND and DIAG(1) on a \(101^2\) grid over \([-6,6]^2\) at eight Gauss--Legendre time nodes.
\end{itemize}

Refining the time quadrature changes the estimated straightness most for the targets with the largest cell stretch. The estimated spread in median Euler error changes little (Sec.~\ref{supp:closed}).

We compare Euler with a reference RK4 integration from common standard Gaussian starts. We compute \(g\) along a subset of those trajectories and compare it with \(S\) and \(A_2=\int_0^1 E_t|a|^2\,dt\) (Sec.~\ref{supp:closed}).

Each target contributes six matched pairs (Tables~\ref{tab:4} and \ref{tab:5}). We test whether the typical matched-pair error differs by at least ten percent and whether the leave-one-target-out residual for \(g\) is less than half the residual for \(S\). Both criteria apply at four and sixteen Euler steps. Sec.~\ref{supp:closed} gives the bootstrap procedure and analyzed target set.

\begin{table}[tbp]
\scriptsize
\setlength{\tabcolsep}{3.5pt}
\caption{Error ratios for couplings with shared target marginals and matched eight-node estimates of \(S\). Ratios compare the reference coupling \(X\) with its MIXRF partner. Entries give geometric means and ranges over targets. The statistic \(Q\) exponentiates the absolute log ratio averaged first within targets and then across targets. Brackets give 95\% target-bootstrap intervals for \(Q\).}
\label{tab:5}
\begin{tabular}{p{\dimexpr 0.124\textwidth-1.488\tabcolsep\relax}p{\dimexpr 0.188\textwidth-2.256\tabcolsep\relax}p{\dimexpr 0.188\textwidth-2.256\tabcolsep\relax}p{\dimexpr 0.188\textwidth-2.256\tabcolsep\relax}p{\dimexpr 0.188\textwidth-2.256\tabcolsep\relax}p{\dimexpr 0.124\textwidth-1.488\tabcolsep\relax}}
\hline
\raggedright reference X & \centering \(N=4\) median ratio & \centering \(N=16\) median ratio & \centering \(N=4\) RMS ratio & \centering \(N=16\) RMS ratio & \centering \(S_X/S_{\mathrm{IND}}\) (range over targets) \tabularnewline
\hline
\raggedright LAM0.50 & \centering 1.07 (0.94 to 1.14) & \centering 1.06 (0.94 to 1.24) & \centering 0.98 (0.41 to 1.25) & \centering 0.89 (0.16 to 1.29) & \centering 0.85 to 0.89 \tabularnewline
\raggedright LAM0.80 & \centering 1.14 (0.90 to 1.45) & \centering 1.11 (0.91 to 1.46) & \centering 0.99 (0.25 to 1.51) & \centering 0.87 (0.09 to 1.54) & \centering 0.75 to 0.81 \tabularnewline
\raggedright LAM1.00 & \centering 1.19 (0.87 to 1.56) & \centering 1.15 (0.89 to 1.57) & \centering 1.01 (0.20 to 1.80) & \centering 0.89 (0.08 to 1.85) & \centering 0.67 to 0.75 \tabularnewline
\raggedright DIAG0.00 & \centering 1.18 (0.91 to 1.62) & \centering 1.12 (0.95 to 1.47) & \centering 1.08 (0.92 to 1.21) & \centering 1.09 (0.92 to 1.21) & \centering 0.19 to 0.35 \tabularnewline
\raggedright DIAG0.50 & \centering 1.05 (0.75 to 1.33) & \centering 1.03 (0.77 to 1.26) & \centering 1.03 (0.93 to 1.10) & \centering 1.05 (0.98 to 1.11) & \centering 0.11 to 0.29 \tabularnewline
\raggedright DIAG0.80 & \centering 1.00 (0.92 to 1.09) & \centering 0.99 (0.90 to 1.06) & \centering 1.01 (0.97 to 1.09) & \centering 1.02 (0.98 to 1.07) & \centering 0.05 to 0.25 \tabularnewline
\raggedright Q (all pairs) & \centering median 1.153 {[}1.123, 1.187{]} & \centering median 1.118 {[}1.089, 1.155{]} & \centering RMS 1.155 {[}1.073, 1.296{]} & \centering RMS 1.199 {[}1.063, 1.473{]} & \centering  \tabularnewline
\hline
\end{tabular}
\end{table}

Matching \(S\) leaves a typical difference of 12 to 15 percent in median error (Table~\ref{tab:5}), with a wider spread in the RMS ratios. A three-component mixture generator gives a similar median-error spread (Sec.~\ref{supp:closed}).

\begin{figure}[tbp]
\centering
\includegraphics[width=\textwidth]{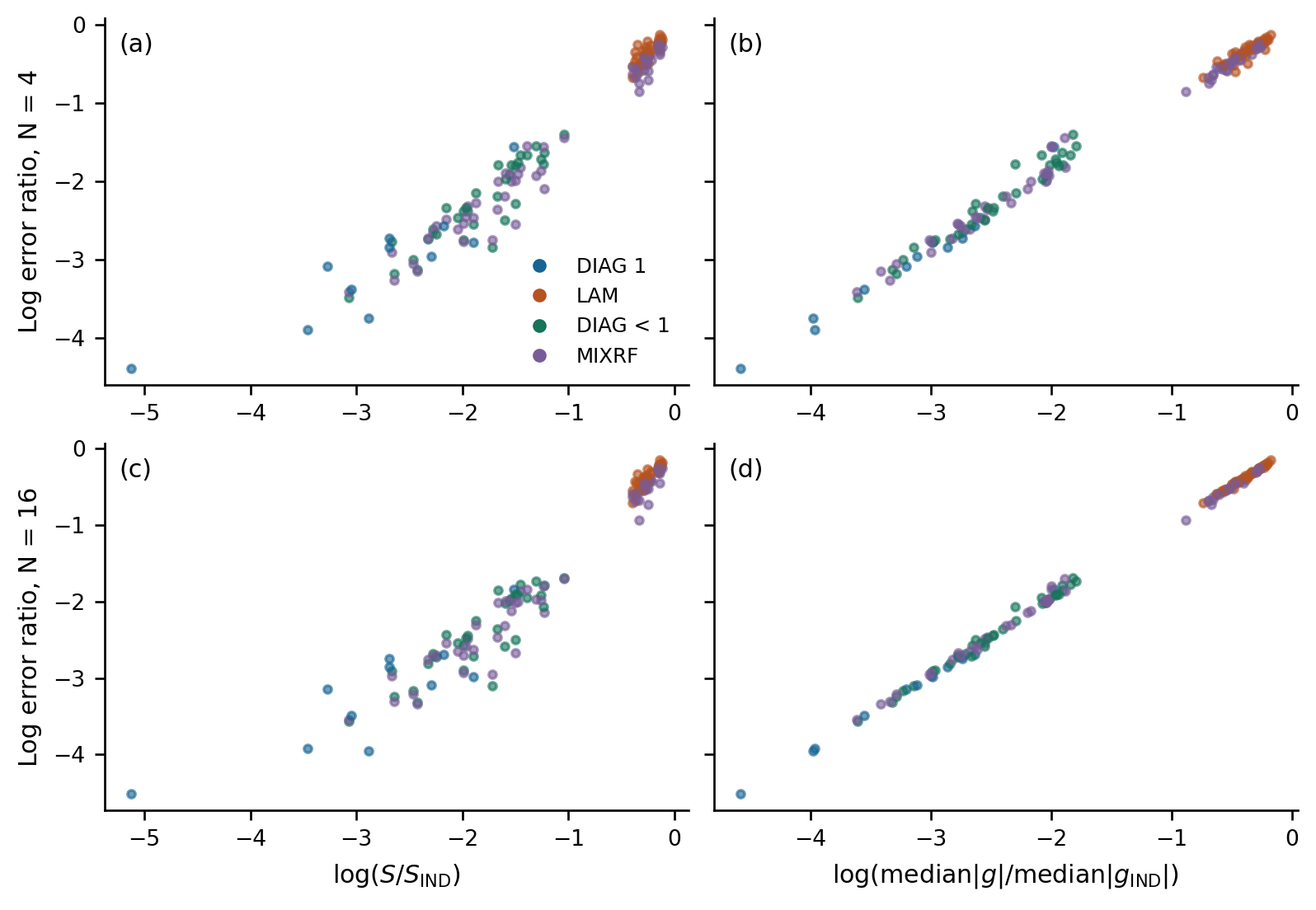}
\caption{Euler error, straightness and the first-order coefficient. (a,c) Median endpoint error versus the eight-node estimate of \(S\), both normalized by the independent coupling of the same target and plotted on log scales. (b,d) The same error versus the similarly normalized median \(|g|\). Rows use four and sixteen steps, respectively. Each panel shows 13 non-independent couplings per target, 11 targets and 800 common starts per coupling.}
\label{fig:1}
\end{figure}

Across the full set of couplings, lower \(S\) generally accompanies smaller median error. The error varies less around the relation with \(g\) (Fig.~\ref{fig:1}).

The held-out log residual for \(g\) is about one seventh of that for \(S\) at sixteen steps, with a smaller difference at four steps (Table~\ref{tab:7}). We compute \(g\) along reference trajectories to test the first-order expansion in Eq.~\eqref{eq:euler-trajectory}. A sampler with half as many steps gives comparable held-out residuals.

\begin{table}[tbp]
\footnotesize
\setlength{\tabcolsep}{3.5pt}
\caption{Leave-one-target-out predictors of the median Euler error (log ratio to the independent coupling of the same target; 13 couplings per target, 11 targets and 800 common starts per coupling). RMS residual pooled over targets, ratio of the g residual to the S residual with a 95 percent bootstrap interval over targets, number of targets with ratio below 1/2, in-sample slopes. The comparison uses the eight-node estimates of \(S\).}
\label{tab:7}
\begin{tabular}{p{\dimexpr 0.052\textwidth-0.832\tabcolsep\relax}p{\dimexpr 0.094\textwidth-1.504\tabcolsep\relax}p{\dimexpr 0.094\textwidth-1.504\tabcolsep\relax}p{\dimexpr 0.158\textwidth-2.528\tabcolsep\relax}p{\dimexpr 0.284\textwidth-4.544\tabcolsep\relax}p{\dimexpr 0.131\textwidth-2.096\tabcolsep\relax}p{\dimexpr 0.094\textwidth-1.504\tabcolsep\relax}p{\dimexpr 0.093\textwidth-1.488\tabcolsep\relax}}
\hline
\raggedright \(N\) & \centering \(S\) & \centering \(g\) & \centering \(\sqrt{A_2}\) & \centering residual ratio [interval] & \centering targets with ratio \(<0.5\) & \centering slope in \(S\) & \centering slope in \(g\) \tabularnewline
\hline
\raggedright 4 & \centering 0.262 & \centering 0.100 & \centering 0.809 & \centering 0.381 {[}0.201, 0.694{]} & \centering 8 of 11 & \centering 1.09 & \centering 0.94 \tabularnewline
\raggedright 16 & \centering 0.295 & \centering 0.041 & \centering 0.861 & \centering 0.140 {[}0.065, 0.244{]} & \centering 11 of 11 & \centering 1.13 & \centering 0.98 \tabularnewline
\hline
\end{tabular}
\end{table}

\begin{figure}[tbp]
\centering
\includegraphics[width=\textwidth]{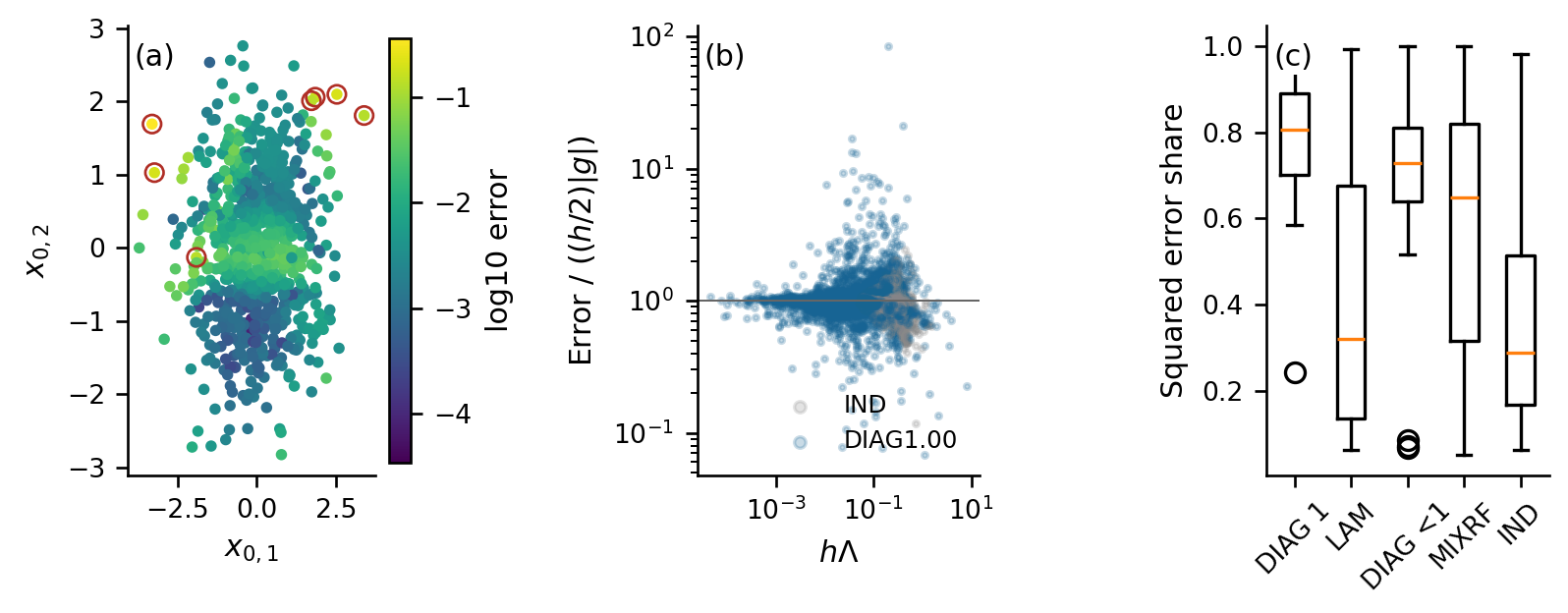}
\caption{Tail contributions to Euler error at \(N=16\). (a) The 800 starting points for the reflow-like pairing on target 21, colored by log error. Circles mark the eight largest errors. (b) Ratio of the error to the first-order prediction versus \(h\Lambda\) for the reflow-like and independent couplings. Here \(h=1/16\) is the Euler step size, and \(\Lambda=\max_t\lambda_{\max}(\operatorname{sym}J(Y_t,t))\) is the largest eigenvalue of the rate-of-strain tensor, maximized over sampled times along each reference trajectory. (c) Share of squared error from the worst 1 percent of 2000 starts, by coupling family.}
\label{fig:3}
\end{figure}

A small fraction of starts accounts for much of the squared error, especially for the reflow-like pairing (Fig.~\ref{fig:3} and Table~\ref{tab:8}). The symmetric part of \(J\) is the rate-of-strain tensor. For the reflow-like pairing, larger peak stretching rates are associated with larger \(|g(1)|\) (Sec.~\ref{supp:closed}). Some tail trajectories depart substantially from the first-order approximation.

\subsection{Closed-form cell-plan models}\label{sec:cellplan}

\begin{figure}[tbp]
\centering
\includegraphics[width=\textwidth]{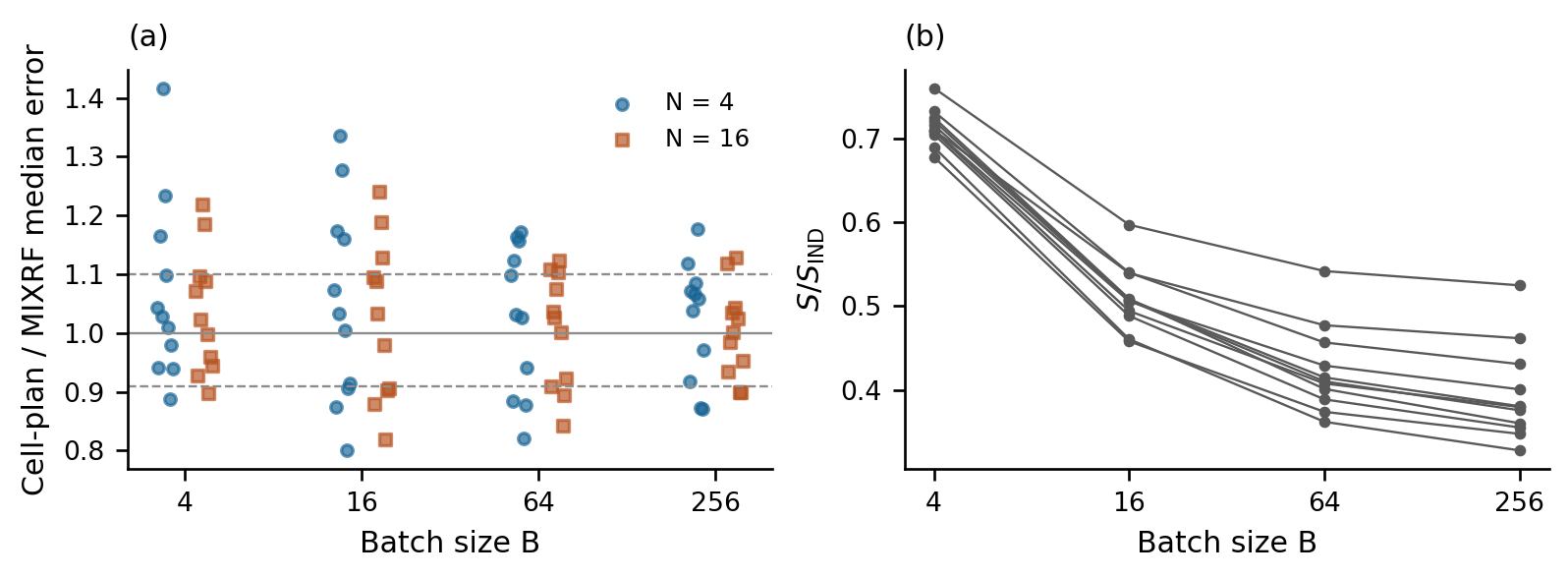}
\caption{Euler error and straightness of cell-plan models. (a) Ratio of median Euler error for a cell-plan model to that of its MIXRF partner, by batch size and step count. Dashed lines indicate a factor of 1.1 above and below parity. (b) \(S/S_{\rm IND}\) of the cell-plan model versus batch size, one line per target.}
\label{fig:4}
\end{figure}

We estimate a plan from the cell-label frequencies of empirical squared-distance assignments and impose a Gaussian coupling within each cell pair. The resulting fields have closed forms (Sec.~\ref{supp:cell}). As batch size increases, the cell-plan model has smaller \(S\) (Fig.~\ref{fig:4}). The typical spread in median error for couplings matched in the quadrature estimate of \(S\) is similar to that in the preceding comparison, with a wider range for individual pairs. The residual advantage of \(g\) is larger at sixteen steps (Table~\ref{tab:9}).

\subsection{Trained fields from true minibatch assignments}\label{sec:mbr}

We train the same multilayer perceptron on the independent, matched mixture and cell-plan couplings, and on actual minibatch assignments. We measure the median distance \(D_N\) between Euler and a reference RK4 integration of the same learned field. Where the exact field is available, we also measure learning error \(L\) against its reference endpoint (Sec.~\ref{supp:trained}).

MB denotes the cell-plan model at \(B=64\). MIX is its quadrature-matched partner from the MIXRF family (Sec.~\ref{sec:closed}). To train MBR, we solve the squared-distance assignment on fresh batches of 64 source and 64 target points and use the resulting pairs. A uniformly selected pair from a batch has the expected batch plan as its law \cite[Definition 3.1]{boite}. IND is the independent baseline.

\begin{figure}[tbp]
\centering
\includegraphics[width=\textwidth]{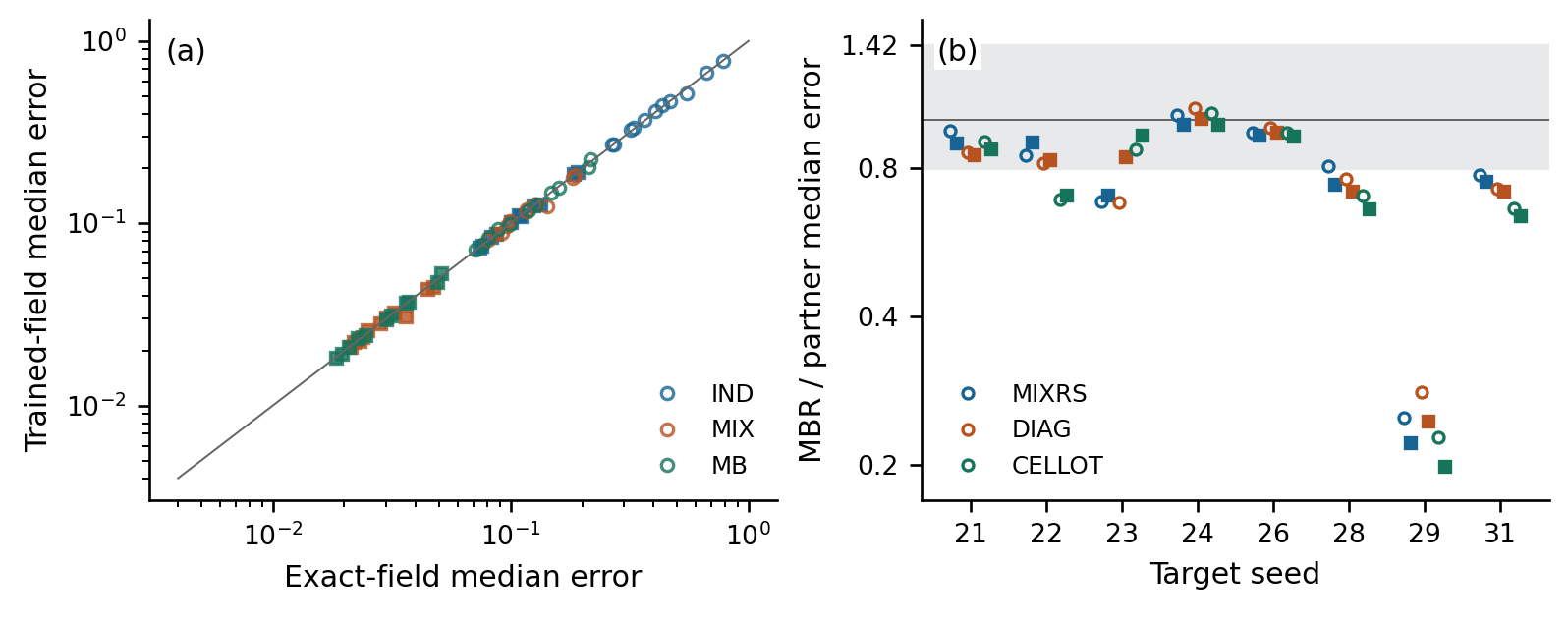}
\caption{Euler error of trained fields. (a) Median error relative to the trained field's RK4 solution versus the median for the corresponding exact field. IND is the independent coupling, MB the cell-plan model and MIX its matched partner. (b) Median error of the field trained on actual minibatch assignments (MBR, \(B=64\)) divided by that of each partner matched to the loss-based estimate of \(S\). The gray band \([0.80,1.42]\) gives the rounded range of observed median-error ratios of cell-plan models to their matched MIXRF partners (Sec.~\ref{sec:cellplan}). In both panels, open circles indicate four steps and filled squares sixteen steps.}
\label{fig:5}
\end{figure}

For the three arms with exact fields, the trained median Euler error agrees with its exact-field counterpart within five percent in 60 of 66 comparisons (Fig.~\ref{fig:5}(a) and Table~\ref{tab:10}). At sixteen steps, learning error is smaller than the discretization error in these runs.

We estimate \(S\) for MBR from the training loss, with a correction calibrated on MB and MIX and checked against IND. This assumes that the learning-error offset transfers to MBR. A finite-sample binning check supports the separation between MBR and MB on selected targets (Sec.~\ref{supp:trained}).

At batch size 64, estimated \(S\) for MBR is roughly one fifth to one quarter of the cell-plan value on ten of eleven targets. The estimates include the sensitivity on target 23 (Tables~\ref{tab:11} and \ref{tab:12}).

The three closed-form partners matched to the loss-based estimate \(S_{\mathrm{MBR}}\) are a member of the MIXRF family (MIXRS), a single Gaussian pair per cell (DIAG), and a non-diagonal entropic cell plan (CELLOT) \cite{r19}. All three partner families yield error ratios below the cell-plan reference band on some targets (Fig.~\ref{fig:5}(b)). Eight targets have partners within the attainable straightness range. The individual ratios appear in Tables~\ref{tab:11}, \ref{tab:14} and \ref{tab:15}.

The first-order coefficient reduces the held-out residual for the trained fields, including MBR, at both step counts. A half-step-count sampler gives a smaller residual at sixteen steps. Sec.~\ref{supp:trained} and Table~\ref{tab:13} report the evaluation and sensitivity checks.

Every matched pair in the closed-form comparison includes MIXRF. The \(S/S_{\rm IND}\) ranges of LAM and DIAG(\(\kappa<1\)) in this design are disjoint (at least 0.67 and at most 0.35, respectively). The cell-plan comparison adds seven LAM partners. The true minibatch comparison uses three partner families on a shared Gaussian lattice, over the eight targets with matches.

\section{Gradient constraints on trained fields}\label{sec:5}

Writing \(v_\theta=\nabla_x\phi_\theta\) imposes a potential-flow inductive bias. The Gaussian score formula~\eqref{eq:gaussian-score} gives a potential field for an independent isotropic Gaussian source. We compare free and gradient-parameterized networks under independent coupling with an isotropic Gaussian, anisotropic Gaussian, or Gaussian-mixture source. We also fit the gradient model directly to the exact field as a supervised comparison and test a soft curl penalty. Sec.~\ref{supp:gradient} describes the training and evaluation.

With an isotropic Gaussian source, curl in the learned field measures one component of learning error. The gradient parameterization reduces velocity error by about 15 percent. A soft curl penalty reduces curl, with overlapping velocity-error ranges across seeds.

For the two cases with rotational exact fields, the hard gradient constraint raises velocity error by roughly six- to sevenfold. The supervised fit for the mixture gives a similar error. These comparisons span training seeds, individual configurations and checkpoints (Tables~\ref{tab:6c}, \ref{tab:6} and \ref{tab:6b}).

\section{Discussion}\label{sec:6}

The virial identity~\eqref{eq:virial} relates coupling straightness to a single radial moment of acceleration. The leading trajectory error depends on how the flow Jacobian propagates local acceleration. Error ratios from fresh minibatch assignments extend below the range observed for the cell-plan mixture comparisons. On target 29, for example, the median error is about 20 to 28 percent of the corresponding median for the three partners matched in estimated straightness.

The strong expansion has numerical support for the studied medians, with larger departures in the tails. The covariance form of the weak expansion was checked on smooth test fields. Its transported-observable integrand has high variance near separatrices in the example of Sec.~\ref{sec:3}. The expansion requires endpoint regularity. With point-mass data the terminal field can be singular.

The error comparisons use constructed targets in two dimensions and one trained architecture. The eight matched targets and their three Gaussian-pair partner families leave the effects of within-cell structure and batch-dependent assignment unresolved. The study measures endpoint error of a sampler relative to a reference for its own field. Distribution-level error across the couplings and image data remain untested.

The potential result concerns independent interpolation. Lim's empirical criterion \cite[Proposition 5.3]{r16} reduces to the antisymmetric part of the posterior-weight derivative when each conditional Jacobian is symmetric. For point-mass targets with a common Gaussian source of covariance \(\sigma^2 I\), the posterior weights satisfy \(\nabla w_i=\frac{t}{(1-t)^2\sigma^2}w_i(x_i-\sum_jw_jx_j)\), so the antisymmetric sum vanishes. The numerical checks include point-mass targets and commuting anisotropic covariances (Table~\ref{tab:1}). Chao et al.~\cite{r12} measure antisymmetric Jacobian components in learned scores. Praneeth et al.~\cite[Sec.~4]{r13} fit an approximate conservative component of a learned velocity for sampler adaptation. Sun et al.~\cite[Eq.~7]{r21} fit a potential with conditional flow matching, and Zhang et al.~\cite[Secs.~3.1--3.2]{r23} obtain the velocity from a potential trained on straight bridges between starts and their own flow endpoints. Hertrich et al.~\cite[Proposition 8]{hertrich} characterize the \(L^2(\rho_t)\) projection onto the closure of gradients of smooth compactly supported potentials. The projected field preserves the marginal path. Tao and Choi's measurements and gradient-model comparisons \cite[Secs.~5.4 and 7]{r6} concern other targets and trained samplers.

\section*{AI-use declaration}
Claude, including Claude Opus 5.5 (Anthropic, through Claude Code), and GPT-6 (OpenAI, through Codex) assisted with literature verification, code development, numerical execution, data analysis and provided editorial assistance with the manuscript under the authors' direction. Deterministic numerical solvers produced the results, and plotting scripts produced the figures. The authors are responsible for the accuracy, originality and final content of this work.

\renewcommand{\bibfont}{\small}

\clearpage
\section*{Supplementary Material}
\suppressfloats[t]
\phantomsection
\label{sec:si}
\newcounter{suppsection}
\renewcommand{\thesuppsection}{S\arabic{suppsection}}
\setcounter{table}{0}
\renewcommand{\thetable}{S\arabic{table}}
\renewcommand{\theHtable}{S\arabic{table}}

\refstepcounter{suppsection}\label{supp:checks}
\subsection*{\thesuppsection\texorpdfstring{\quad}{ } Numerical checks of the governing identities}

For mixtures of jointly Gaussian endpoint pairs, we evaluate the velocity as the posterior-weighted sum of component velocities \cite[Theorem 5, Eq.~6]{hertrich}. We evaluate \(\rho\) and \(\Sigma\) in closed form and use float64 automatic differentiation for \(J\), \(\partial_tv\) and \(\nabla\!\cdot(\rho\Sigma)\). Three routes do not use that algebra: (a) data as point masses with a Gaussian source, where the posterior over data points is a softmax; (b) direct quadrature over \(x_0\) of the defining integral using only the densities \(p_0,p_1\) (scipy densities, trapezoid on a 0.06 grid) with central finite differences of step \(2\times10^{-3}\), including a smooth ring density as non-Gaussian data and an isotropic Student-\(t\)(5) source; (c) a one-dimensional coupling of three overlapping deterministic streams \(x_1=T_k(x_0)\) with three different nonlinear maps, inverted by Newton iteration. A Monte Carlo of conditional paths with local-quadratic kernel regression (10 batches of \(10^6\) pairs, bandwidths 0.30 and 0.20) is bias-limited and supports only the field definitions at the 5 to 15 percent level where the field varies fast; it is not used as a precision test.

The curl-free prediction holds on all three routes for the isotropic Gaussian source (curl ratio at rounding level, \(4\times10^{-6}\) for the finite-difference quadrature including ring data) and the measured curl is nonzero for an anisotropic Gaussian, a mixture source and an isotropic non-Gaussian (Student-\(t\)) source, so isotropy alone does not suffice. It also vanishes for the commuting Gaussian covariances tested, so ``rotational for an anisotropic source'' is generic and not universal. The two-point-mass tests support necessity in \(d=2\) and \(d=3\) against six non-Gaussian or anisotropic sources. The Jeans residual is at rounding level on the closed-form route, the point-mass route and the nonlinear three-stream coupling, and to the finite-difference level on the quadrature route; its residual for 31 comparison couplings (independent, correlated, cell-level optimal transport, cell-smoothed reflow, mixtures) is between \(3\times10^{-16}\) and \(2\times10^{-14}\), and for the 44 couplings of Section 4 at most \(1.2\times10^{-11}\). The Gaussian schedule formula~\eqref{eq:gaussian-schedule} holds for the linear, cosine and quadratic schedules to \(2\times10^{-15}\); with an anisotropic source the score formula holds to the same level while the Jacobian asymmetry is 0.74 to 0.78.

For the strong Euler law on the independent coupling (target seed 5, 2000 points) the pointwise residual of the strong law relative to the prediction is 0.40, 0.20 and 0.081 at \(N=64,256,1024\) (the residual decreases with step size), the RMS error times \(N\) approaches 2.7 to 2.8 (half the RMS of \(g\), 5.76), and the median pointwise residual is 0.3 percent at \(N=256\); the RK4 reference is self-converged to \(5\times10^{-11}\). The mean square is dominated by a few trajectories near separatrices, so \(N\gtrsim256\) is needed before the first-order law holds for the RMS. The weak law in the \(g\) form gives measured over predicted 0.992, 0.998, 1.002 for \(\psi=|x|^2\) at \(N=128,512,2048\) (first target; 20000 points) and 0.94 to 1.12 for the other two couplings, within the sampling error of the coefficient (1 to 4 percent). The \(\Sigma:\nabla^2u\) form follows by integration by parts in the Jeans balance~\eqref{eq:jeans} (checked on smooth test fields), but its Monte Carlo estimate has a heavy-tailed integrand near separatrices, where \(\nabla^2(\psi\circ\Phi_{t\to1})\) can become large: it agrees with the \(g\) form at \(t=0.8\) (\(5.87\pm0.06\) against \(5.88\pm0.13\)) and carries no information at \(t\le0.5\). The weak law is therefore supported in the \(g\) form, and in the \(\Sigma\) form only on smooth test fields.

\refstepcounter{suppsection}\label{supp:closed}
\subsection*{\thesuppsection\texorpdfstring{\quad}{ } Closed-form comparisons on Gaussian mixtures}

The first generator draws the means of five components from \(N(0,1.5^2I)\). For both generators each covariance is \(0.35^2 AA^T+0.05 I\), with \(A\) a standard Gaussian matrix. The mixture weights follow a symmetric Dirichlet distribution with concentration 5 per component.

Per coupling: \(S\), \(A_2\), the Euler error of each of 2000 common starting points at \(N=2,4,8,16,32\) against a 200-step RK4 solution, and for the first 800 starts \(g(1)\), the propagator of \(\dot F=JF\) and \(\max_t\lambda_{\max}(\mathrm{sym}\,J)\). The starts are shared standard Gaussian draws. Checks: the marginals at \(t=0\) and \(t=1\) agree with the nominal marginal densities at the probed locations to \(2\times10^{-7}\) (source, lattice accuracy) and, for the mixtures, \(1.9\times10^{-12}\) (target, at the 30 heaviest components of two targets; the mixture drops components of weight below \(10^{-13}\), total mass \(2\times10^{-12}\); a probe at the means of the lowest-weight components gives relative errors up to \(8\times10^{-2}\) for four members, where the density is below \(10^{-13}\) of the peak); the Jeans balance~\eqref{eq:jeans} holds for 44 couplings of the four families to \(1.2\times10^{-11}\). Here \(A_2=\int_0^1 E_t|a|^2\,dt\).

We use the lower limit of the 95 percent target-level bootstrap interval for \(Q\), the exponential of the mean absolute log ratio of median errors, to compare the typical spread with a benchmark of 1.10 at \(N=4\) and \(N=16\). For held-out predictions, the comparison benchmarks a residual ratio \(g/S\) below one half on at least 80 percent of targets, with a bootstrap upper limit below one half. The analysis uses the first 11 of 20 prespecified target seeds. The subset was fixed before evaluating the comparison statistics, after individual results were available.

The eight-node quadrature used for matching gives relative residuals \(\lvert\Delta S/S\rvert\le 2.1\times10^{-9}\). On the first target set, re-evaluating the same 66 pairs at 16 time nodes on the same spatial grid gives a median absolute mismatch of 0.01 percent and a maximum of 2.67 percent, on target 23. For targets 23 and 25, 32 nodes give \(S_{\mathrm{ref}}/S_{\mathrm{IND}}\) of 0.171 and 0.224, respectively, compared with 0.149 and 0.220 at eight nodes. For target 27 the ratio is 0.094 at 16 nodes and 0.101 at eight. The reflow-like pairing on target 23 still changes by about 2 percent in \(S\) between 16 and 32 nodes.

To estimate the effect of matching at 16 time nodes, we interpolate the log median error as a function of log \(S\) within the evaluated MIXRF family for each target. The estimated \(Q_{\rm median}\) is 1.152 at four steps and 1.117 at sixteen, compared with 1.153 and 1.118 for the eight-node matches. The interpolation uses the existing trajectory errors. Replacing the eight-node estimates of \(S\) with the 16-node estimates in the held-out fits gives residual ratios of 0.36 and 0.13, respectively, with the same numbers of targets below one half.

For the eight-node quadrature matches, \(Q_{\rm median}=1.153\) {[}1.123, 1.187{]} at \(N=4\) and 1.118 {[}1.089, 1.155{]} at \(N=16\). The sixteen-step lower bound falls below the benchmark.

The reference coupling's error exceeds the mixture's in 76 percent of the pairs (mean log ratio 0.124 and 0.098 for LAM at \(N=4\) and 16, 0.070 and 0.044 for DIAG), so it measures a systematic difference from the mixture partner as well as scatter. For these quadrature-matched pairs, the median Euler error of two couplings of these families differs by a typical 12 to 15 percent (per-target values 6 to 26 percent; 95 percent interval of the mean 9 to 19 percent), with single pairs from 0.75 to 1.62 (4 steps) and 0.77 to 1.57 (16 steps); the starting-point interval excludes 1 for 91 and 92 percent of the pairs, an assessment of finite-start variability conditional on these targets. The largest mean difference is LAM against the mixture (geometric mean 1.13 and 1.10 for LAM over MIXRF, 1.07 and 1.04 for DIAG over MIXRF).

The RMS error is less well matched: \(Q_{\rm rms}=1.155\) {[}1.073, 1.296{]} and 1.199 {[}1.063, 1.473{]}, with single ratios from 0.20 to 1.80 (\(N=4\)) and 0.075 to 1.85 (\(N=16\)), the extreme values coming from the target with the largest cell stretch (seed 23, stretch 342).

Across the whole range \(S\) falls by a factor 4.5 to 170 from the independent to the reflow-like coupling (S\(_{\rm ref}\)/S\(_{\rm IND}\) from 0.006 to 0.22) while the median error at \(N=16\) falls by a factor 6.6 to 97 (5.3 to 85 at \(N=4\); Fig.~\ref{fig:1}). A power law in \(S\) alone describes this with slope 1.09 (\(N=4\)) and 1.13 (\(N=16\)) and an RMS log residual of 0.24 and 0.27, that is a factor 1.3 in the error (the matched-pair spread above is 1.12 to 1.15).

On the first 800 common starts, the leave-one-target-out RMS log residual of the median error is 0.100 for \(g\) and 0.262 for \(S\) at \(N=4\) (ratio 0.38, interval 0.20 to 0.69) and 0.041 against 0.295 at \(N=16\) (ratio 0.14, interval 0.065 to 0.24). Eight targets are below the residual-ratio threshold at four steps and all eleven at sixteen steps. The four-step advantage is less consistent across targets.

The median error over the first 800 starts (those that carry \(g\)) equals \((h/2)\,\mathrm{median}|g|\) with ratio 0.925 to 1.251 at \(N=16\) and 0.958 to 1.105 at \(N=32\) over all 154 couplings (median 0.998 and 1.000), with no fitted parameter. This verifies the first-order trajectory error formula~\eqref{eq:euler-trajectory} again on couplings with the constructed target; it is not evidence that \(g\) is a practical predictor.

A sampler with half as many steps gives held-out residuals comparable to those of \(g\), which we evaluate along reference trajectories.

A few starts dominate the RMS error, especially in the straighter couplings (Fig.~\ref{fig:3} and Table~\ref{tab:8}). In this tail analysis, the worst 1 percent of 2000 starts carries a median 0.29 of the squared error at \(N=16\) for the independent coupling and 0.81 for the reflow-like pairing (RMS over median 2.1 and 14.5). The worst 1 percent have median \(|x_0|=2.8\) to 2.9 (class medians) against 1.18 for all starts, and the 4.4 percent of starts with \(|x_0|>2.5\) carry a median 0.84 of the squared error for the reflow-like pairing (0.31 for the independent coupling and LAM); the tail is the far Gaussian tail of the source, where the cell maps are most stretched (the median RMS error of a target at \(N=16\) ranges from 0.037 to 3.7 and rank-correlates with its maximum cell stretch, Spearman 0.74 over 11 targets), which is partly a property of this construction of the target. The Spearman correlation between the error and \(|g|\) over 800 starts is 0.99 or 1.00 (class medians; lowest 0.93), and for the worst 8 starts the ratio of the error to \((h/2)|g|\) at \(N=16\) has class medians 0.87 to 0.96. Over all 800 starts the class medians of the share with a ratio outside \([0.5,2]\) are at most 0.9 percent, but in 33 of the 154 couplings (all among the straighter ones: 12 DIAG(\(\kappa<1\)), 15 MIXRF, 5 reflow-like, 1 LAM) more than 1 percent of the starts lie outside, up to 17 percent, and in 24 of those 33 couplings more starts exceed 2 than fall below 0.5, with equal shares in one coupling, so the first-order law does not hold for every tail start. The heavy tail of the error is therefore largely the heavy tail of \(|g|\) (which correlates with the flow-map stretch, Spearman 0.62, and with \(\max_t\lambda_{\max}\), 0.74, for the reflow-like pairing), with a further minority of starts beyond the first-order regime in the straightest couplings. The single worst start has a ratio as low as 0.21 and as high as 2.7, and the RMS-level law \(\mathrm{RMS}\,e/((h/2)\,\mathrm{RMS}|g|)\) ranges from 0.22 to 1.00 at \(N=16\) (median 0.90) and 0.36 to 1.08 at \(N=32\) (median 0.95), which suggests that a handful of starts with very large \(|g|\) are outside the first-order regime. On target 23, a Richardson estimate from Euler endpoint errors agrees with the trajectory-based median \(|g|\) within 2 percent but can differ substantially for individual tail starts.

\subsubsection*{Three-component mixture targets}

Six targets use three-component mixtures, with component means drawn from \(N(0,1.2^2I)\), covariance scale 0.35, and seeds 51 to 56. Each target has 14 couplings (84 total). The analysis applies the same rules to these six targets separately. For these targets, \(Q_{\rm median}=1.131\) {[}1.078, 1.182{]} at \(N=4\) and 1.098 {[}1.058, 1.136{]} at \(N=16\), with both lower bounds below 1.10; \(Q_{\rm rms}=1.073\) and 1.055; the residual ratio \(g/S\) is 0.321 {[}0.204, 0.564{]} at \(N=4\) (4 of 6 targets below 1/2, 5 needed) and 0.075 {[}0.044, 0.128{]} at \(N=16\) (6 of 6), with leave-one-target-out residuals 0.225 and 0.270 for \(S\) and 0.072 and 0.020 for \(g\), with more consistent gains at sixteen steps, as for the five-component mixtures; the first-order ratio of the median error is 0.93 to 1.06 at \(N=16\) and 0.94 to 1.02 at \(N=32\); the class medians of the share of the squared error in the worst 1 percent are 0.33 (independent), 0.37 (LAM), 0.82 (DIAG(\(\kappa<1\))), 0.76 (MIXRF) and 0.94 (reflow-like), and the Spearman correlation of error and \(|g|\) is at least 0.96 in every coupling.

\refstepcounter{suppsection}\label{supp:cell}
\subsection*{\thesuppsection\texorpdfstring{\quad}{ } Cell-plan construction and comparisons}

In the cell-plan comparison we build a closed-form model from cell-label frequencies of empirical assignments \cite{r3,r4}: \(B\) independent source points (cell label, then a point in the cell) and \(B\) independent target points (component label, then a point in the component) are paired by the assignment that minimises \(\sum|x_0-x_1|^2\); the frequencies of the label pairs over \(4\times10^5\) pairs estimate the plan \(\Gamma_B(i,j)\), which is projected onto the exact cell weights by iterative proportional fitting (marginal error at most \(2\times10^{-8}\); the diagonal pairing is added at weight \(10^{-6}c_i\) so that cells of tiny weight have a feasible support, and carries at most \(1.6\times10^{-4}\) of the mass); inside a cell pair \(\mathrm{Cov}(x_0,x_1)=\kappa_B s^2 D\phi_j^T\), with \(\kappa_B\) (0.08 to 0.44) fitted to the in-pair dependence measured on the same minibatches. On a separate target used to assess numerical sensitivity, \(4\times10^5\) against \(1.6\times10^6\) pairs changes \(S\) and the median errors by at most 0.4 percent; replacing the fitted in-cell dependence by independence raises \(S\) by 2 percent at \(B=4\) and 12 percent at \(B=256\), so the results concern this representation, which does not reproduce the straightness of minibatch transport at \(B\ge16\) (see the block on trained fields below).

On the 11 targets of the first target set, \(B=4,16,64,256\) give \(S/S_{\rm IND}=0.68\) to 0.76, 0.46 to 0.60, 0.36 to 0.54 and 0.33 to 0.52 (Fig.~\ref{fig:4}(b)); each is paired with the mixture of the same \(S\) (44 pairs) and, where \(S\) is in range, with LAM (7 pairs, \(B=4\)); DIAG(\(\kappa\)) never reaches these \(S\). The fields were evaluated by a second implementation (torch, float64, a different algebra), which reproduces the closed-form reference of four couplings to \(5\times10^{-13}\).

For the cell-plan comparisons (Table~\ref{tab:9}), \(Q_{\rm median}=1.121\) {[}1.094, 1.148{]} at \(N=4\) and 1.097 {[}1.075, 1.116{]} at \(N=16\), with both lower bounds below 1.10. The cell-plan coupling has a slightly larger median error than its quadrature-matched mixture at small \(B\) (geometric mean 1.06 at \(B=4\), \(N=4\)) and the same at large \(B\) (1.00 at \(N=16\) for \(B\ge64\)), with single pairs from 0.80 to 1.42, and a 13 percent smaller one than its matched LAM partner. The RMS errors are not matched at all (ratio 0.04 to 3.8), with the mixture carrying the heavy-tailed reflow-like part. The ratio of the leave-one-target-out residuals of \(g\) and \(S\) is 0.40 {[}0.29, 0.49{]} at \(N=4\) (8 of 11 targets below 1/2) and 0.11 {[}0.08, 0.15{]} at \(N=16\) (11 of 11), with more consistent gains at sixteen steps. The first-order law holds for every cell-plan coupling tested (median error over \((h/2)\,\mathrm{median}|g|\) 0.96 to 1.05 at \(N=16\), 0.98 to 1.02 at \(N=32\)). The median-error spread factors are about 1.12 and 1.10, with wider spreads for individual pairs and RMS error.

\refstepcounter{suppsection}\label{supp:trained}
\subsection*{\thesuppsection\texorpdfstring{\quad}{ } Trained fields and minibatch comparisons}

To compare actual minibatch assignments with the cell-plan model and other partners, we train the same MLP (three hidden layers of 128 SiLU units, inputs \((x,t)\), 4096 fresh pairs per step, Adam with a cosine schedule from \(2.02\times10^{-3}\) toward \(2\times10^{-5}\), 20000 steps, float32, and \(t\sim U(0,1)\) for a 200-step reference solve within the training interval) on couplings of the 11 targets of Table~\ref{tab:4}, and measure on the 2000 starting points of the first target set: \(D_N\), the median over starts of \(|\mathrm{Euler}_N(v_\theta)-\mathrm{RK4}_{200}(v_\theta)|\) (the discretization error of the sampler for its own field); \(L\), the median of \(|\mathrm{RK4}_{200}(v_\theta)-X_e|\) (learning error of the transport map, \(X_e\) the endpoint of the exact field); and \(T_N\), the median of \(|\mathrm{Euler}_N(v_\theta)-X_e|\).

Four arms: IND; MIX, the mixture of the reflow-like pairing and IND with the \(S\) of the cell-plan coupling at \(B=64\) (MB); MB itself; and MBR, real minibatch optimal transport (64 source points from \(N(0,I)\) and 64 target points from \(p_1\), paired by the assignment problem on \(|x_0-x_1|^2\), with 64 fresh groups of 64 pairs at each training step), whose exact field is not known.

We use four targets with the smallest cell stretch (maximum stretch 8 to 11) with three training seeds, the other seven with two seeds for IND, MIX and MB and one seed for MBR. This gives 97 runs across the four arms and 44 runs for the mixture partners below. The training comparisons on the four lowest-stretch targets informed the criteria for the other seven. The partner reference interval is based on the cell-plan results.

\subsubsection*{Trained and exact fields}

The trained \(D_N\) equals the exact-field median error to within 5 percent in 60 of 66 (target, coupling, \(N\)) cells (range 0.85 to 1.06, median 1.00; the six outside are on targets 22, 23 and 25). For the matched pair, \(\rho_N\), the geometric mean over training seeds of \(D_N(\mathrm{MB})/D_N(\mathrm{MIX})\), is 0.82 to 1.27 on all 11 targets, within the empirical reference interval 0.80 to 1.42 from the cell-plan mixture partners (the LAM partners reach 0.758), and equals the exact-field ratio of the cell-plan comparison medians to a median of 1.2 percent (trained over exact 0.96 to 1.17; the largest ratio occurs on target 23 at \(N=16\)). The ratio of \(D_N\) for IND over MB is 2.99 to 4.33, within 5 percent of the exact ratio in all 22 cells. The seed-to-seed spread of \(D_N\) for MB and MIX (maximum over minimum, \(N=4\) and 16) is at most 1.14. Table~\ref{tab:10} and Fig.~\ref{fig:5}(a) give the cell-plan comparisons.

The learning-to-discretization error ratio \(L/D_{16}\) has median 0.09 (IND), 0.26 (MIX) and 0.24 (MB) over the 11 targets and is largest, 0.55, for MIX on target 23 (maximum stretch 342); \(T_{16}/D_{16}\) is 0.96 to 1.03 (IND), 0.89 to 1.16 (MIX) and 1.03 to 1.13 (MB). At 16 steps the median error against the exact flow is close to the median discretization error of the learned field in these runs, with \(L\) smaller than \(D_{16}\); the relative velocity error of the MIX models is 0.019 to 0.039. This concerns these 11 targets, this MLP and 20000 steps.

\subsubsection*{Estimating minibatch straightness}

By the loss decomposition~\eqref{eq:loss-decomposition}, the population loss of a fixed network is \(S\) plus the learning error energy. The offset \(\delta\) is estimated per target from the MB and MIX runs, whose \(S\) is evaluated by quadrature (\(\delta=-0.002\) to \(+0.096\); its standard deviation over the runs is 0.0003 to 0.0067 on eight targets, 0.023 on target 27, 0.108 on 25 and 0.158 on 23, where the training loss is heavy-tailed), and the MBR estimate is \(S_{\rm MBR}=\overline{\rm loss}-\delta\), using the mean of the last 200 stochastic training losses. Calibration on MB and MIX estimates the offset for MBR. It gives no deterministic bound on the MBR estimate. The estimator is checked on IND, which is held out of \(\delta\): the predicted \(S_{\rm IND}\) is within 0.13 percent of the quadrature value on the four initial targets, within 0.33 percent on five of the seven additional targets and 2.9 percent (target 27) and 1.8 percent (target 23) on the other two (column ``IND check'' of Table~\ref{tab:11}).

With fresh minibatch assignments at \(B=64\), \(S_{\rm MBR}\) is 0.19 to 0.26 of the cell-plan value on 10 of the 11 targets (0.447 on target 23, whose estimate is unresolved to about 16 percent), and 0.076 to 0.104 of \(S_{\rm IND}\) on those 10 against 0.36 to 0.54 for the cell-plan coupling (Table~\ref{tab:11}). On three targets (30, 25, 27) \(S_{\rm MBR}\) is below the \(S\) of the reflow-like pairing (0.382 against 0.427, 0.719 against 1.524, 0.416 against 0.478): the real coupling at \(B=64\) is straighter than the lowest member of the mixture family. A sweep over \(B\) on target 24 (one training seed) shows the gap opening with \(B\): ratio 0.81 at \(B=4\), 0.48 at 16, 0.22 at 64 (Table~\ref{tab:12}).

An exploratory check without a learned field uses the binned conditional variance of \(u\) given \(x_t\) from \(2\times10^7\) real minibatch pairs per \(B\) with bins of width 0.1 (the corresponding population quantity bounds \(S\) from above, while finite sampling, spatial truncation, dropped sparse bins and time quadrature affect its estimate) and gives 2.302, 0.960 and 0.361 at \(B=4\), 16 and 64 on target 24 (0.06 to 0.33 percent from the loss route; the same estimator gives 3.986 for the independent coupling, exact 4.003), 0.270 and 0.293 on targets 31 and 29 (loss route 0.272 and 0.293), and estimates 0.97 (25), 0.69 (27) and 2.87 (23) on three heavy-tailed targets, where the loss route is flagged (0.72, 0.42, 0.99); for target 23 the binned estimate exceeds the cell-plan value of 2.22. The empirical training loss is 1.05, with heavy-tailed fluctuations, so the exact \(S\) remains unresolved.

The plan-frequency and within-cell sensitivity checks assess the precision of the cell-plan estimate. The difference in straightness between the cell plan and fresh assignments remains across targets (Table~\ref{tab:11}).

\subsubsection*{Partners matched to estimated straightness}

The mixture partners in Fig.~\ref{fig:5}(b) combine the reflow-like pairing and IND. We match the quadrature estimate of the mixture's \(S\) to \(f\,S_{\rm MBR}\) by bisection, then train with three seeds on the four initial targets and two seeds on the four additional targets with partners. The prespecified reference window is 0.80 to 1.42 at both step counts. On the four initial targets, \(\rho_N\) is 1.025 and 0.981 (target 24, \(N=4\), 16), 0.773 and 0.752 (31), 0.808 and 0.744 (28) and 0.950 and 0.901 (21); varying the \(S\) estimate by 5 percent moves \(\rho_N\) to 0.791 to 1.046 (\(f=0.95\)) and 0.704 to 1.006 (\(f=1.05\)). Both sensitivity settings include ratios outside the reference window. At \(f=1\), re-evaluating the DIAG and MIXRS partners at 16 time nodes places their \(S\) within 0.03 percent of the loss-based target value on seven targets and about 9 to 10 percent above it on target 23.

On the additional targets with a partner (22, 29, 26, 23) \(\rho_N\) is 0.848 and 0.906, 0.250 and 0.223, 0.943 and 0.934, 0.684 and 0.705; the \(S\) estimate of target 23 is unresolved, and target 29, where the reflow-like pairing is extremely straight (\(S_{\rm ref}=0.017\)), has the largest gap (a factor 4.5 in \(D_{16}\)). Targets 29 and 23 fall below the reference window at both step counts. Targets 30, 25 and 27 are outside the attainable straightness range of the mixture family.

Over the eight targets with a partner the geometric mean of \(\rho_N\) is 0.73 (\(N=4\)) and 0.71 (\(N=16\)) and the typical factor \(Q\) (the exponential of the mean absolute log ratio, as in the closed-form comparison) is 1.37 and 1.41; without target 29 it is 1.18 and 1.19, and without 29 and 23 it is 1.14 and 1.16, against 1.12 and 1.10 for the cell-plan couplings. The ratio of the same MBR runs to MIX of the cell-plan \(S\) (different estimated straightness) is 0.13 to 0.23 on the four initial targets. The error ratio \(\rho_N\) is 0.68 to 1.03 on seven of the eight targets (0.74 to 1.03 without the unresolved target 23) and 0.22 to 0.25 on target 29. The full comparison uses all available partners. The mixture has a straight bulk, and the single-mode partner uses one Gaussian pair per cell.

\subsubsection*{First-order Euler error of trained fields}

For each of the 97 trained models, \(g\) is integrated along the RK4 trajectory of the learned field (100 steps, \(g'=Jg+a\) with \(J=\partial v_\theta/\partial x\) and \(a=\partial_tv_\theta+Jv_\theta\) by forward-mode derivatives of the network) from the first 800 common Gaussian starts, also used for the Euler errors in this comparison; applying the same numerical procedure to the exact field of the cell-plan coupling of target 24 reproduces the closed-form reference \(g\) to \(5\times10^{-14}\). The first-order law holds for the trained fields: \(D_N/((h/2)\,\mathrm{median}|g|)\) has median 0.995 (0.94 to 1.32) at \(N=16\) and 0.997 (0.97 to 1.19) at \(N=32\) over the 97 runs, and 1.03 and 1.02 over the 19 MBR runs, whose largest value (target 23) falls from 1.32 to 1.19 between \(N=16\) and 32.

With the rows MIX, MB and MBR per target and the trained IND of the same target as baseline, the same evaluation as for the closed-form fields gives leave-one-target-out RMS log residuals of 0.330 for \(S\) and 0.078 for \(g\) at \(N=4\) (ratio 0.236 {[}0.127, 0.330{]}, 11 of 11 targets below 1/2) and 0.352 and 0.049 at \(N=16\) (ratio 0.139 {[}0.070, 0.237{]}, 11 of 11).

At the MBR row alone the residual of \(S\) is 0.53 and 0.56 and that of \(g\) 0.11 and 0.08 (ratio 0.20 {[}0.09, 0.30{]} and 0.14 {[}0.06, 0.26{]}): the line in \(S\) through the closed-form couplings misplaces the real procedure in both directions, by factors (observed over predicted error) from 0.26 (target 29) to 1.9 (target 26) at \(N=16\), while \(g\) places it within 0.09 in log on ten targets and 0.23 on target 23, whose \(S\) estimate is unresolved (Table~\ref{tab:13}).

The held-out slope in \(\log g\) is 0.92 to 0.98, near the value 1 of the first-order law; that in \(\log S\) is 1.03 to 1.20. Training seed 0 alone, and \(S_{\rm MBR}\) scaled by 0.9 or 1.1, leave the ratios at 0.231 to 0.241 (\(N=4\)) and 0.135 to 0.142 (\(N=16\)) in an exploratory sensitivity analysis. As for exact fields, a sampler run at half the steps predicts the trained-field error worse than \(g\) at \(N=4\) (residual 0.134 with \(N=2\), against 0.078 for \(g\)) and better at \(N=16\) (0.023 with \(N=8\), against 0.049), so this comparison evaluates the first-order law along reference trajectories. The first-order contrast thus extends to the trained fields in this benchmark.

\subsubsection*{Single-mode and entropic partners}

The mixture partner is a global mixture of the reflow-like pairing and IND. The comparison uses DIAG(\(\kappa\)) of the first target set instead: one Gaussian pair per cell (source cell \(i\) to target component \(i\), \(\mathrm{Cov}(x_0,x_1)=\kappa s^2D\phi_i^T\)), whose \(S\) decreases monotonically from IND-like at \(\kappa=0\) to \(S_{\rm ref}\) at \(\kappa=1\) (checked on a 21-point scan per target), with \(\kappa^*\) bisected to \(S_{\rm MBR}\) (0.52 to 0.95; the same eight targets have a partner). The models are trained exactly as for the matched mixture (20 runs). The error ratio \(\rho_N\) is 0.68 to 1.05 on seven targets and 0.28 and 0.25 on target 29; outside the window are 31, 28 and 29 (both \(N\)) and 23 (\(N=4\)). The typical factor is \(Q=1.40\) and 1.39 (1.23 and 1.19 without target 29), and the geometric mean of \(\rho_N\) is 0.72 and 0.72 (Table~\ref{tab:14}). The two partners agree with each other: at the same nominal \(S\) the trained error of DIAG(\(\kappa^*\)) is 0.84 to 1.10 times that of the mixture (typical factor 1.06, 16 values). The two closed-form families, both based on diagonal pairs, give errors within about 6 percent for the nominal matches, and real minibatch optimal transport has a smaller error than both on seven of eight targets (all but 24; against DIAG(\(\kappa^*\)) below 0.80 on 31, 28 and 29 at both \(N\)). Checks: the trained error of DIAG(\(\kappa^*\)) is within 5 percent of its exact-field median error in 12 of 16 cells (0.89 to 1.11); the converged loss over the quadrature \(S\) is 0.993 to 1.005 in 18 of 20 runs (1.25 and 0.976 in the two runs on the heavy-tailed target 23), a calibration check on couplings outside the calibration; the ratio of median \(|g|\) of the trained fields predicts \(\rho_N\) to 0.80 to 1.02 (median 0.97). 

The mixture and single-mode partners use the diagonal cell pairing. This comparison uses CELLOT(\(\epsilon\)) instead: entropic optimal transport \cite{r19} with cost \(|\mu_i-\phi_j|^2\) between the source cells \(N(\mu_i,s^2I)\) and the target components \(N(\phi_j,B_j)\) of the first target set, with the Gaussian optimal-transport map \(x_1=\phi_j+B_j^{1/2}(x_0-\mu_i)/s\) inside each pair (\(C=sB_j^{1/2}\)). Its design uses the common lattice source and common target for every \(\epsilon\) (Sinkhorn marginal errors below \(10^{-13}\), at most \(7\times10^{-11}\) after dropping plan entries below \(10^{-14}\)), its plan has off-diagonal mass. The posterior weights of the overlapping source cells vary with \(x_0\). Given the sampled source label \(i\), the plan fixes the probabilities of the target labels. The map inside each pair is monotone. \(S\) increases monotonically with \(\epsilon\) on every scan; at \(\epsilon^*\) matched to \(S_{\rm MBR}\) (\(\epsilon^*/s^2\) 0.45 to 1.87) the plan keeps 0.15 to 0.67 of its mass on the diagonal (22831 to 76211 pair components), and all eight targets have a partner.

Training is the same as for DIAG (20 runs). The error ratio \(\rho_N\) is 0.64 to 1.03 on seven targets and 0.23 and 0.20 on target 29; outside the window are 31, 28, 22 and 29 (both \(N\)); \(Q=1.446\) and 1.475 (1.23 and 1.24 without target 29), geometric mean 0.70 and 0.68 (Table~\ref{tab:15}).

Without any network, the exact-field median error of CELLOT(\(\epsilon^*\)) over that of DIAG(\(\kappa^*\)) at the same nominal \(S\) is 0.76 to 1.25 (geometric mean 1.03 at \(N=4\), 1.06 at \(N=16\); first 800 starts), and 1.06 to 1.25 on the four targets where real minibatch transport is furthest below the partners. On those four targets the non-diagonal plan increases the median error relative to DIAG. The three closed-form families give comparable median errors across this set (pairwise trained-error ratios 0.76 to 1.28), and real minibatch optimal transport has a smaller error than all three on seven of eight targets (all but 24; by 4 to 36 percent on six, by a factor of 3.5 to 5 on target 29). Checks: trained over exact median error 0.99 to 1.05 (16 of 16 within 5 percent); loss over quadrature \(S\) 0.994 to 1.009 in 18 of 20 runs (1.24 and 1.15 on the heavy-tailed target 23), a further calibration check; \(g_{\rm MBR}/g_{\rm CELLOT}\) of the trained fields predicts \(\rho_N\) to 0.81 to 0.99 (median 0.96).

The \(\epsilon/s^2\) scan begins at 0.158 and ends at the first point with \(S>S_{\rm MBR}\). The lower bound restricts the comparison to the range where Sinkhorn solves meet the marginal tolerance. The quadrature match uses a tolerance of \(10^{-6}\), and the exact-field comparisons use 800 common starts. The loss-based \(S_{\rm MBR}\) estimate has 0.5 to 3 percent noise. All three partners are Gaussian-pair mixtures on the same cell lattice, so the comparison does not resolve structure below the cell scale or reproduce the data-dependent pairing within each minibatch.

\refstepcounter{suppsection}\label{supp:gradient}
\subsection*{\thesuppsection\texorpdfstring{\quad}{ } Gradient-constrained models}

We use two-dimensional problems with a five-component Gaussian mixture target (seed 11) and three sources: isotropic \(N(0,I)\) (exact curl 0), anisotropic \(\mathrm{diag}(2,0.3)\) and a three-component Gaussian mixture (data to data; exact curl-to-gradient ratio 0.409 and 0.375 for the last two). Independent coupling, linear interpolant, multilayer perceptrons with three hidden layers of 128 SiLU units and inputs \((x,t)\), flow-matching loss on 4096 fresh pairs per step, 20000 Adam steps with a cosine schedule, float32. Models: FREE (two outputs), GRAD (\(v=\nabla_x\phi_\theta(x,t)\), conservative by construction, same loss), GRADFLOOR (the same gradient network regressed directly on the exact field \(v^*\), providing a supervised comparison in the same parametrization), and FREE with a penalty \(\lambda E[\mathrm{curl}^2]\), \(\lambda=1\), on 512 points per step. Evaluation: density-weighted over a grid at \(t=0.1,0.3,0.5,0.7,0.9\) against the exact field and Jacobian (closed form, autodiff). For the individual configurations in Table~\ref{tab:6}, we fit one FREE model per source with a second seed for the isotropic and mixture sources. The other configurations in that table use one run, so differences below a few percent between models are not resolved (the two isotropic seeds differ by 2 to 3 percent in the error and vorticity columns and by 9 percent in the antisymmetric share).

With the independent isotropic Gaussian source, the exact field is curl-free by the Gaussian score formula~\eqref{eq:gaussian-score}, so the vorticity of the fitted field measures one component of learning error. For the unconstrained model, the antisymmetric part accounts for 5.6 to 6.1 percent of the Jacobian error energy. Its evolution during training is shown in Table~\ref{tab:6b}. The supervised gradient fit has velocity error 0.0048, compared with 0.0155 for the gradient model trained with flow matching. The comparison leaves the contributions of optimization and regression-target noise unresolved.

The seed-averaged comparison uses three seeds per isotropic and mixture configuration, three for the anisotropic FREE model, two for the anisotropic GRAD model and two for each penalised configuration (Table~\ref{tab:6c}). For the isotropic source the FREE model has curl/\(|J|=0.041\) to \(0.043\) over three seeds, the penalty (weight 1, two seeds) gives 0.0077 to 0.0083 and leaves the velocity error at 0.0184 against 0.0187 without it (seed ranges 0.0179 to 0.0190 and 0.0182 to 0.0194, overlapping), and the conservative parametrisation lowers the velocity error by 15 percent (ratio of means 0.854; over all pairs of seeds 0.80 to 0.93) and the Jacobian error by 10 percent (0.110 against 0.123).

For the rotational exact fields the GRAD model's velocity error is 5.64 times the FREE model's for the mixture source (pairs 5.57 to 5.75) and 7.24 times for the anisotropic source (7.06 to 7.58), near the supervised fit error of 0.135 for the mixture; the penalised FREE model on the mixture source is 5.16 times the unpenalised one (two seeds, 0.1243 against 0.0241). Seed-averaged results and individual configurations appear in Tables~\ref{tab:6c} and \ref{tab:6}. The comparison uses one architecture, one target and one interpolation schedule.
The source controls and residuals for Sec.~\ref{sec:3} are given in Tables~\ref{tab:1}--\ref{tab:3}. Target characteristics and endpoint error distributions appear in Tables~\ref{tab:4} and \ref{tab:8}.
For Lim's empirical criterion (Sec.~\ref{sec:6}), point-mass targets in dimensions two and three give median relative residuals \(1.4\times10^{-11}\) and \(3.2\times10^{-11}\) at the finite-difference level. The corresponding ranges are 0.10 to 0.20 with per-point variances and 0.14 to 0.22 with an anisotropic source.
\begin{table}[htp]
\small
\setlength{\tabcolsep}{3.5pt}
\caption{RMS curl over RMS \(|\nabla v|\) of the exact field, by source (independent coupling; Gaussian-mixture target unless stated; density-weighted)}
\label{tab:1}
\begin{tabular}{p{\dimexpr 0.355\textwidth-2.840\tabcolsep\relax}p{\dimexpr 0.237\textwidth-1.896\tabcolsep\relax}p{\dimexpr 0.172\textwidth-1.376\tabcolsep\relax}p{\dimexpr 0.236\textwidth-1.888\tabcolsep\relax}}
\hline
\raggedright source & \centering closed form (4 times, min to max) & \centering direct quadrature (3 times x 6 points, rms) & \centering point-mass data (3 times) \tabularnewline
\hline
\raggedright isotropic Gaussian & \centering 1.5e-16 to 5.0e-16 & \centering 4.21e-06 & \centering 1.8e-16 to 2.3e-16 \tabularnewline
\raggedright isotropic Gaussian, variance 2.89 / 2.25 & \centering 1.0e-16 to 3.3e-16 & \centering - & \centering 1.2e-16 to 2.1e-16 \tabularnewline
\raggedright anisotropic Gaussian diag(2,0.3) / diag(1.5,0.4) & \centering 2.0e-01 to 5.3e-01 & \centering 3.12e-01 & \centering 4.7e-01 to 5.0e-01 \tabularnewline
\raggedright Gaussian mixture source (data to data) & \centering 1.7e-01 to 4.2e-01 & \centering 4.34e-01 & \centering - \tabularnewline
\raggedright anisotropic source, commuting Gaussian target & \centering 0.0e+00 to 0.0e+00 & \centering - & \centering - \tabularnewline
\raggedright anisotropic source, non-commuting Gaussian target & \centering 1.4e-01 to 7.1e-01 & \centering - & \centering - \tabularnewline
\raggedright isotropic Student-t(5) source & \centering - & \centering 1.89e-01 & \centering - \tabularnewline
\raggedright isotropic Gaussian source, ring data (non-Gaussian data) & \centering - & \centering 3.73e-06 & \centering - \tabularnewline
\hline
\end{tabular}
\end{table}
\begin{table}[htp]
\small
\setlength{\tabcolsep}{3.5pt}
\caption{Relative residuals of the identities on four routes}
\label{tab:2}
\begin{tabular}{p{\dimexpr 0.335\textwidth-3.350\tabcolsep\relax}p{\dimexpr 0.178\textwidth-1.780\tabcolsep\relax}p{\dimexpr 0.162\textwidth-1.620\tabcolsep\relax}p{\dimexpr 0.178\textwidth-1.780\tabcolsep\relax}p{\dimexpr 0.147\textwidth-1.470\tabcolsep\relax}}
\hline
\raggedright identity & \centering closed-form fields & \centering point-mass data & \centering quadrature + finite differences & \centering 1D nonlinear three-stream coupling \tabularnewline
\hline
\raggedright Jeans balance \eqref{eq:jeans}, divided by \(\rho\) & \centering max 1.8e-15 & \centering max 3.9e-15 & \centering max 2.5e-03 (FD) & \centering max 3.3e-15 \tabularnewline
\raggedright Virial identity \eqref{eq:virial} & \centering max 3.4e-03 (grid truncation) & \centering max 2.3e-03 & \centering - & \centering max 1.2e-10 \tabularnewline
\raggedright Tested Jeans balance \eqref{eq:tested-jeans} & \centering max 5.7e-03 & \centering - & \centering - & \centering - \tabularnewline
\raggedright Isotropic Gaussian covariance \eqref{eq:conditional-covariance} & \centering max 1.4e-15 & \centering max 1.5e-15 & \centering - & \centering - \tabularnewline
\raggedright Gaussian score formula \eqref{eq:gaussian-score} & \centering max 1.5e-15 & \centering - & \centering - & \centering - \tabularnewline
\hline
\end{tabular}
\end{table}
\begin{table}[htp]
\small
\setlength{\tabcolsep}{3.5pt}
\caption{Antisymmetric part of the exact Jacobian for two-point-mass data, \(\|J-J^T\|/\|J\|\), median over six random data sets}
\label{tab:3}
\begin{tabular}{p{\dimexpr 0.596\textwidth-3.576\tabcolsep\relax}p{\dimexpr 0.202\textwidth-1.212\tabcolsep\relax}p{\dimexpr 0.202\textwidth-1.212\tabcolsep\relax}}
\hline
\raggedright source & \centering \(d=2\) & \centering \(d=3\) \tabularnewline
\hline
\raggedright isotropic Gaussian & \centering 6.8e-16 & \centering 6.1e-16 \tabularnewline
\raggedright anisotropic Gaussian & \centering 5.2e-01 & \centering 6.7e-01 \tabularnewline
\raggedright isotropic Student-t(5) & \centering 1.2e-01 & \centering 2.6e-01 \tabularnewline
\raggedright isotropic sech & \centering 4.9e-01 & \centering 3.9e-01 \tabularnewline
\raggedright isotropic smoothed Laplace & \centering 3.1e-01 & \centering 3.7e-01 \tabularnewline
\raggedright isotropic quartic exp(-r\^{}2/2 - 0.1 r\^{}4) & \centering 5.0e-02 & \centering 5.9e-03 \tabularnewline
\raggedright isotropic scale mixture of Gaussians & \centering 5.9e-02 & \centering 2.1e-01 \tabularnewline
\hline
\end{tabular}
\end{table}
\begin{table}[htp]
\footnotesize
\setlength{\tabcolsep}{3.5pt}
\caption{Constructed targets and baseline Euler errors. The straightness columns give the eight-node quadrature estimates for the independent coupling \(S_{\mathrm{IND}}\) and the reflow-like pairing \(S_{\mathrm{ref}}\). Sec.~\ref{supp:closed} reports estimates at finer time resolution. Maximum stretch is the largest singular value of the cell map. Euler errors use \(N=16\) and 2000 starts.}
\label{tab:4}
\begin{tabular}{p{\dimexpr 0.069\textwidth-1.242\tabcolsep\relax}p{\dimexpr 0.113\textwidth-2.034\tabcolsep\relax}p{\dimexpr 0.113\textwidth-2.034\tabcolsep\relax}p{\dimexpr 0.120\textwidth-2.160\tabcolsep\relax}p{\dimexpr 0.103\textwidth-1.854\tabcolsep\relax}p{\dimexpr 0.103\textwidth-1.854\tabcolsep\relax}p{\dimexpr 0.103\textwidth-1.854\tabcolsep\relax}p{\dimexpr 0.103\textwidth-1.854\tabcolsep\relax}p{\dimexpr 0.173\textwidth-3.114\tabcolsep\relax}}
\hline
\raggedright seed & \centering \(S_{\mathrm{IND}}\) & \centering \(S_{\mathrm{ref}}/S_{\mathrm{IND}}\) & \centering max stretch & \centering IND median & \centering IND RMS & \centering ref. median & \centering ref. RMS & \centering share of total squared error with \(\|x_0\|>2.5\) (ref.) \tabularnewline
\hline
\raggedright 21 & \centering 4.30 & \centering 0.038 & \centering 11 & \centering 0.1099 & \centering 0.1714 & \centering 0.0045 & \centering 0.0308 & \centering 0.75 \tabularnewline
\raggedright 22 & \centering 4.26 & \centering 0.056 & \centering 23 & \centering 0.1007 & \centering 0.2806 & \centering 0.0020 & \centering 0.0597 & \centering 0.84 \tabularnewline
\raggedright 23 & \centering 4.66 & \centering 0.149 & \centering 342 & \centering 0.1333 & \centering 0.3126 & \centering 0.0067 & \centering 4.6984 & \centering 1.00 \tabularnewline
\raggedright 24 & \centering 4.00 & \centering 0.068 & \centering 8 & \centering 0.0831 & \centering 0.1514 & \centering 0.0047 & \centering 0.0473 & \centering 0.90 \tabularnewline
\raggedright 25 & \centering 6.92 & \centering 0.220 & \centering 61 & \centering 0.1838 & \centering 1.0785 & \centering 0.0279 & \centering 0.9043 & \centering 0.45 \tabularnewline
\raggedright 26 & \centering 6.09 & \centering 0.068 & \centering 31 & \centering 0.1916 & \centering 0.3949 & \centering 0.0117 & \centering 0.0794 & \centering 0.55 \tabularnewline
\raggedright 27 & \centering 4.73 & \centering 0.101 & \centering 166 & \centering 0.1248 & \centering 1.1175 & \centering 0.0057 & \centering 0.7885 & \centering 1.00 \tabularnewline
\raggedright 28 & \centering 3.73 & \centering 0.031 & \centering 10 & \centering 0.1078 & \centering 0.1417 & \centering 0.0023 & \centering 0.0330 & \centering 0.97 \tabularnewline
\raggedright 29 & \centering 2.84 & \centering 0.006 & \centering 25 & \centering 0.0756 & \centering 0.1125 & \centering 0.0008 & \centering 0.0013 & \centering 0.26 \tabularnewline
\raggedright 30 & \centering 3.76 & \centering 0.114 & \centering 13 & \centering 0.0873 & \centering 0.2718 & \centering 0.0059 & \centering 0.1446 & \centering 0.84 \tabularnewline
\raggedright 31 & \centering 2.81 & \centering 0.047 & \centering 9 & \centering 0.0738 & \centering 0.1184 & \centering 0.0023 & \centering 0.0241 & \centering 0.73 \tabularnewline
\hline
\end{tabular}
\end{table}
\begin{table}[htp]
\scriptsize
\setlength{\tabcolsep}{3.5pt}
\caption{Cell-plan models derived from minibatch label frequencies against partners matched in the quadrature estimate of \(S\), 11 targets, 51 matched pairs per \(N\): matched-\(S\) spread (rule: lower bound at least 1.10), signed ratio of the median error of the cell-plan model to that of the matched mixture (geometric mean over targets) for \(B\) = 4, 16, 64, 256, leave-one-target-out RMS log residuals of the median error for \(S\) and \(g\), and their ratio (rule: below 1/2 in at least 9 of 11 targets, upper bound below 1/2).}
\label{tab:9}
\begin{tabular}{p{\dimexpr 0.037\textwidth-0.518\tabcolsep\relax}p{\dimexpr 0.201\textwidth-2.814\tabcolsep\relax}p{\dimexpr 0.256\textwidth-3.584\tabcolsep\relax}p{\dimexpr 0.106\textwidth-1.484\tabcolsep\relax}p{\dimexpr 0.106\textwidth-1.484\tabcolsep\relax}p{\dimexpr 0.201\textwidth-2.814\tabcolsep\relax}p{\dimexpr 0.093\textwidth-1.302\tabcolsep\relax}}
\hline
\raggedright N & \centering \(Q_{\rm median}\) [95\% CI] & \centering cell plan / mixture, \(B=4,16,64,256\) & \centering residual for \(S\) & \centering residual for \(g\) & \centering residual ratio [95\% CI] & \centering targets below 1/2 \tabularnewline
\hline
\raggedright 4 & \centering 1.121 {[}1.094, 1.148{]} & \centering 1.058, 1.037, 1.019, 1.018 & \centering 0.142 & \centering 0.057 & \centering 0.402 {[}0.289, 0.487{]} & \centering 8 of 11 \tabularnewline
\raggedright 16 & \centering 1.097 {[}1.075, 1.116{]} & \centering 1.033, 1.016, 0.999, 0.999 & \centering 0.151 & \centering 0.016 & \centering 0.107 {[}0.083, 0.149{]} & \centering 11 of 11 \tabularnewline
\hline
\end{tabular}
\end{table}

\begin{table}[htp]
\scriptsize
\setlength{\tabcolsep}{3.5pt}
\caption{Heavy tail of the Euler error at N = 16 by coupling class (median over couplings of the per-coupling statistic; minimum to maximum in brackets): share of total squared error carried by the worst 1 percent of 2000 starts, RMS over median, ratio of measured to first-order error for the eight largest errors among 800 starts, median \(\|x_0\|\) of the worst 1 percent (all starts: 1.18), share of total squared error from starts with \(\|x_0\|>2.5\) (prior probability 0.044).}
\label{tab:8}
\begin{tabular}{p{\dimexpr 0.130\textwidth-1.820\tabcolsep\relax}p{\dimexpr 0.106\textwidth-1.484\tabcolsep\relax}p{\dimexpr 0.154\textwidth-2.156\tabcolsep\relax}p{\dimexpr 0.146\textwidth-2.044\tabcolsep\relax}p{\dimexpr 0.154\textwidth-2.156\tabcolsep\relax}p{\dimexpr 0.154\textwidth-2.156\tabcolsep\relax}p{\dimexpr 0.156\textwidth-2.184\tabcolsep\relax}}
\hline
\raggedright class & \centering couplings & \centering share, worst 1 percent & \centering RMS / median & \centering first-order ratio, worst 8 & \centering median \(\|x_0\|\), worst 1 percent & \centering share, \(\|x_0\|>2.5\) \tabularnewline
\hline
\raggedright independent & \centering 11 & \centering 0.29 {[}0.06, 0.98{]} & \centering 2.1 {[}1.3, 9.0{]} & \centering 0.93 {[}0.67, 1.07{]} & \centering 2.83 {[}1.38, 3.12{]} & \centering 0.31 {[}0.16, 0.98{]} \tabularnewline
\raggedright LAM & \centering 33 & \centering 0.32 {[}0.07, 0.99{]} & \centering 2.3 {[}1.3, 13.6{]} & \centering 0.96 {[}0.71, 1.46{]} & \centering 2.87 {[}1.29, 3.14{]} & \centering 0.31 {[}0.12, 0.99{]} \tabularnewline
\raggedright DIAG kappa \textless{} 1 & \centering 33 & \centering 0.73 {[}0.07, 1.00{]} & \centering 8.1 {[}1.6, 727.2{]} & \centering 0.87 {[}0.61, 1.00{]} & \centering 2.89 {[}2.40, 3.06{]} & \centering 0.77 {[}0.17, 1.00{]} \tabularnewline
\raggedright MIXRF & \centering 66 & \centering 0.65 {[}0.05, 1.00{]} & \centering 4.7 {[}1.3, 643.5{]} & \centering 0.92 {[}0.66, 1.27{]} & \centering 2.85 {[}1.38, 3.12{]} & \centering 0.70 {[}0.13, 1.00{]} \tabularnewline
\raggedright reflow-like pairing (DIAG 1) & \centering 11 & \centering 0.81 {[}0.24, 1.00{]} & \centering 14.5 {[}1.7, 698.7{]} & \centering 0.91 {[}0.68, 1.26{]} & \centering 2.90 {[}2.30, 3.03{]} & \centering 0.84 {[}0.26, 1.00{]} \tabularnewline
\hline
\end{tabular}
\end{table}
\begin{table}[htp]
\scriptsize
\setlength{\tabcolsep}{3.5pt}
\caption{Trained against exact fields. Ratio of the median Euler error of MB to that of MIX (trained over exact value, geometric mean over training seeds; reference values from the closed-form cell-plan fields) and of IND to MB, at \(N=4\) and 16; the median trained \(D_{16}\) over the exact one, and \(L/D_{16}\), per coupling. Targets by increasing cell stretch.}
\label{tab:10}
\begin{tabular}{p{\dimexpr 0.052\textwidth-0.832\tabcolsep\relax}p{\dimexpr 0.091\textwidth-1.456\tabcolsep\relax}p{\dimexpr 0.135\textwidth-2.160\tabcolsep\relax}p{\dimexpr 0.135\textwidth-2.160\tabcolsep\relax}p{\dimexpr 0.118\textwidth-1.888\tabcolsep\relax}p{\dimexpr 0.118\textwidth-1.888\tabcolsep\relax}p{\dimexpr 0.189\textwidth-3.024\tabcolsep\relax}p{\dimexpr 0.162\textwidth-2.592\tabcolsep\relax}}
\hline
\raggedright seed & \centering max stretch & \centering MB/MIX, N = 4 (trained / exact) & \centering MB/MIX, N = 16 & \centering IND/MB, N = 4 & \centering IND/MB, N = 16 & \centering \(D_{16}\) trained over exact (IND, MIX, MB) & \centering \(L/D_{16}\) (IND, MIX, MB) \tabularnewline
\hline
\raggedright 24 & \centering 8 & \centering 1.032 / 1.032 & \centering 1.023 / 1.026 & \centering 3.37 / 3.30 & \centering 3.46 / 3.40 & \centering 1.014, 1.000, 0.997 & \centering 0.10, 0.26, 0.25 \tabularnewline
\raggedright 31 & \centering 9 & \centering 0.932 / 0.942 & \centering 0.920 / 0.922 & \centering 3.54 / 3.51 & \centering 3.82 / 3.77 & \centering 1.002, 0.991, 0.989 & \centering 0.09, 0.18, 0.22 \tabularnewline
\raggedright 28 & \centering 10 & \centering 0.984 / 1.026 & \centering 0.969 / 1.002 & \centering 3.53 / 3.41 & \centering 3.73 / 3.60 & \centering 1.024, 1.022, 0.989 & \centering 0.07, 0.17, 0.18 \tabularnewline
\raggedright 21 & \centering 11 & \centering 1.085 / 1.098 & \centering 1.102 / 1.108 & \centering 3.48 / 3.40 & \centering 3.54 / 3.50 & \centering 1.002, 0.996, 0.990 & \centering 0.07, 0.22, 0.20 \tabularnewline
\raggedright 30 & \centering 13 & \centering 0.935 / 0.878 & \centering 0.924 / 0.894 & \centering 3.92 / 3.97 & \centering 4.15 / 4.20 & \centering 0.999, 0.977, 1.010 & \centering 0.10, 0.28, 0.29 \tabularnewline
\raggedright 22 & \centering 23 & \centering 0.911 / 0.884 & \centering 0.904 / 0.910 & \centering 3.97 / 4.16 & \centering 4.33 / 4.42 & \centering 1.005, 1.032, 1.026 & \centering 0.09, 0.26, 0.24 \tabularnewline
\raggedright 29 & \centering 25 & \centering 0.820 / 0.821 & \centering 0.828 / 0.842 & \centering 3.77 / 3.81 & \centering 4.07 / 4.08 & \centering 0.993, 1.013, 0.995 & \centering 0.07, 0.13, 0.19 \tabularnewline
\raggedright 26 & \centering 31 & \centering 1.158 / 1.156 & \centering 1.092 / 1.103 & \centering 3.79 / 3.68 & \centering 4.01 / 3.90 & \centering 0.996, 0.979, 0.969 & \centering 0.07, 0.21, 0.20 \tabularnewline
\raggedright 25 & \centering 61 & \centering 1.219 / 1.163 & \centering 1.184 / 1.074 & \centering 2.99 / 3.09 & \centering 3.52 / 3.60 & \centering 1.020, 0.947, 1.043 & \centering 0.09, 0.31, 0.24 \tabularnewline
\raggedright 27 & \centering 166 & \centering 1.158 / 1.172 & \centering 1.131 / 1.123 & \centering 3.20 / 3.18 & \centering 3.39 / 3.44 & \centering 1.000, 1.008, 1.015 & \centering 0.09, 0.31, 0.28 \tabularnewline
\raggedright 23 & \centering 342 & \centering 1.267 / 1.123 & \centering 1.212 / 1.036 & \centering 3.27 / 3.44 & \centering 3.40 / 3.57 & \centering 0.950, 0.850, 0.995 & \centering 0.11, 0.55, 0.31 \tabularnewline
\hline
\end{tabular}
\end{table}
\begin{table}[htp]
\scriptsize
\setlength{\tabcolsep}{3.5pt}
\caption{\(S\) of real minibatch optimal transport at \(B=64\), estimated from the training loss (MBR; three training seeds on targets 24, 31, 28, 21 and one on the others), against the cell-plan coupling of the same \(B\); the check of the estimator on IND; and the ratio of the trained discretization error of MBR to that of the mixture partner MIXRS matched to the loss-based estimate of \(S\) (\(f=1.00\), geometric mean over the partner's training seeds; three seeds on targets 24, 31, 28 and 21, and two on the other targets with partners). ``No partner'': \(S_{\rm MBR}\) is below the minimum \(S\) of the mixture family.}
\label{tab:11}
\begin{tabular}{p{\dimexpr 0.058\textwidth-1.044\tabcolsep\relax}p{\dimexpr 0.101\textwidth-1.818\tabcolsep\relax}p{\dimexpr 0.106\textwidth-1.908\tabcolsep\relax}p{\dimexpr 0.158\textwidth-2.844\tabcolsep\relax}p{\dimexpr 0.072\textwidth-1.296\tabcolsep\relax}p{\dimexpr 0.181\textwidth-3.258\tabcolsep\relax}p{\dimexpr 0.086\textwidth-1.548\tabcolsep\relax}p{\dimexpr 0.119\textwidth-2.142\tabcolsep\relax}p{\dimexpr 0.119\textwidth-2.142\tabcolsep\relax}}
\hline
\raggedright seed & \centering max stretch & \centering \(S_{\rm MBR}\) & \centering \(S_{\rm cell-plan}\) & \centering ratio & \centering \(S_{\rm MBR}/S_{\rm IND}\) & \centering IND check & \centering real / mixture, N = 4 & \centering N = 16 \tabularnewline
\hline
\raggedright 24 & \centering 8 & \centering 0.3599 & \centering 1.6321 & \centering 0.221 & \centering 0.090 & \centering -0.13\% & \centering 1.025 & \centering 0.981 \tabularnewline
\raggedright 31 & \centering 9 & \centering 0.2716 & \centering 1.1677 & \centering 0.233 & \centering 0.096 & \centering +0.01\% & \centering 0.773 & \centering 0.752 \tabularnewline
\raggedright 28 & \centering 10 & \centering 0.3482 & \centering 1.5307 & \centering 0.227 & \centering 0.093 & \centering -0.03\% & \centering 0.808 & \centering 0.744 \tabularnewline
\raggedright 21 & \centering 11 & \centering 0.3655 & \centering 1.5561 & \centering 0.235 & \centering 0.085 & \centering -0.06\% & \centering 0.950 & \centering 0.901 \tabularnewline
\raggedright 30 & \centering 13 & \centering 0.3825 & \centering 1.7152 & \centering 0.223 & \centering 0.102 & \centering -0.03\% & \centering no partner & \centering no partner \tabularnewline
\raggedright 22 & \centering 23 & \centering 0.4044 & \centering 1.6554 & \centering 0.244 & \centering 0.095 & \centering -0.33\% & \centering 0.848 & \centering 0.906 \tabularnewline
\raggedright 29 & \centering 25 & \centering 0.2926 & \centering 1.1414 & \centering 0.256 & \centering 0.103 & \centering -0.11\% & \centering 0.250 & \centering 0.223 \tabularnewline
\raggedright 26 & \centering 31 & \centering 0.4622 & \centering 2.2779 & \centering 0.203 & \centering 0.076 & \centering -0.30\% & \centering 0.943 & \centering 0.934 \tabularnewline
\raggedright 25 & \centering 61 & \centering 0.7188 & \centering 3.7506 & \centering 0.192 & \centering 0.104 & \centering -0.25\% & \centering no partner & \centering no partner \tabularnewline
\raggedright 27 & \centering 166 & \centering 0.4161 & \centering 2.0300 & \centering 0.205 & \centering 0.088 & \centering -2.89\% & \centering no partner & \centering no partner \tabularnewline
\raggedright 23 & \centering 342 & \centering 0.9947 & \centering 2.2247 & \centering 0.447 & \centering 0.213 & \centering -1.81\% & \centering 0.684 & \centering 0.705 \tabularnewline
\hline
\end{tabular}
\end{table}

\begin{table}[htp]
\footnotesize
\setlength{\tabcolsep}{3.5pt}
\caption{Real minibatch optimal transport against the cell-plan coupling as a function of the batch size (target 24).}
\label{tab:12}
\begin{tabular}{p{\dimexpr 0.049\textwidth-0.686\tabcolsep\relax}p{\dimexpr 0.115\textwidth-1.610\tabcolsep\relax}p{\dimexpr 0.160\textwidth-2.240\tabcolsep\relax}p{\dimexpr 0.089\textwidth-1.246\tabcolsep\relax}p{\dimexpr 0.132\textwidth-1.848\tabcolsep\relax}p{\dimexpr 0.221\textwidth-3.094\tabcolsep\relax}p{\dimexpr 0.234\textwidth-3.276\tabcolsep\relax}}
\hline
\raggedright B & \centering \(S\) real (loss minus delta) & \centering \(S\) cell-plan & \centering ratio & \centering \(S\) real over \(S_{\rm IND}\) & \centering \(D_4\) real & \centering \(D_{16}\) real \tabularnewline
\hline
\raggedright 4 & \centering 2.301 & \centering 2.835 & \centering 0.812 & \centering 0.575 & \centering 0.1478 & \centering 0.03773 \tabularnewline
\raggedright 16 & \centering 0.958 & \centering 1.978 & \centering 0.484 & \centering 0.239 & \centering 0.0587 & \centering 0.01440 \tabularnewline
\raggedright 64 & \centering 0.360 & \centering 1.632 & \centering 0.221 & \centering 0.090 & \centering 0.0220 (3 seeds) & \centering 0.00511 (3 seeds) \tabularnewline
\hline
\end{tabular}
\end{table}
\begin{table}[htp]
\footnotesize
\setlength{\tabcolsep}{3.5pt}
\caption{Real minibatch optimal transport (\(B=64\)) against the single-mode partner DIAG(\(\kappa^*\)) matched to the loss-based estimate of \(S\): \(\kappa^*\); seeds of the partner and of MBR; ratio \(\rho_N\) of the trained median Euler errors at \(N=4\) and 16 (geometric means over training seeds); the ratio against the mixture partner of Table~\ref{tab:11} (MIXRS, \(N=4\), 16); trained over exact median error of the partner (\(N=4\), 16); and the first-order prediction \(g_{\rm MBR}/g_{\rm DIAG}\) over the observed ratio (\(N=4\), 16; first 800 starts).}
\label{tab:14}
\begin{tabular}{p{\dimexpr 0.060\textwidth-0.960\tabcolsep\relax}p{\dimexpr 0.100\textwidth-1.600\tabcolsep\relax}p{\dimexpr 0.100\textwidth-1.600\tabcolsep\relax}p{\dimexpr 0.100\textwidth-1.600\tabcolsep\relax}p{\dimexpr 0.100\textwidth-1.600\tabcolsep\relax}p{\dimexpr 0.180\textwidth-2.880\tabcolsep\relax}p{\dimexpr 0.180\textwidth-2.880\tabcolsep\relax}p{\dimexpr 0.180\textwidth-2.880\tabcolsep\relax}}
\hline
\raggedright tgt & \centering \(\kappa^*\) & \centering seeds & \centering \(\rho_4\) & \centering \(\rho_{16}\) & \centering MIXRS & \centering tr./ex. & \centering g/obs. \tabularnewline
\hline
\raggedright 24 & \centering 0.873 & \centering 3/3 & \centering 1.055 & \centering 1.009 & \centering 1.02/0.98 & \centering 1.01/1.01 & \centering 0.96/0.98 \tabularnewline
\raggedright 31 & \centering 0.737 & \centering 3/3 & \centering 0.726 & \centering 0.721 & \centering 0.77/0.75 & \centering 0.99/1.00 & \centering 1.02/1.00 \tabularnewline
\raggedright 28 & \centering 0.686 & \centering 3/3 & \centering 0.761 & \centering 0.718 & \centering 0.81/0.74 & \centering 0.96/0.97 & \centering 0.93/0.99 \tabularnewline
\raggedright 21 & \centering 0.700 & \centering 3/3 & \centering 0.861 & \centering 0.854 & \centering 0.95/0.90 & \centering 0.97/0.97 & \centering 1.01/0.99 \tabularnewline
\raggedright 22 & \centering 0.761 & \centering 2/1 & \centering 0.818 & \centering 0.833 & \centering 0.85/0.91 & \centering 1.04/1.06 & \centering 0.97/0.98 \tabularnewline
\raggedright 29 & \centering 0.516 & \centering 2/1 & \centering 0.282 & \centering 0.246 & \centering 0.25/0.22 & \centering 1.01/1.01 & \centering 0.83/0.98 \tabularnewline
\raggedright 26 & \centering 0.947 & \centering 2/1 & \centering 0.963 & \centering 0.949 & \centering 0.94/0.93 & \centering 1.05/1.11 & \centering 0.89/0.92 \tabularnewline
\raggedright 23 & \centering 0.563 & \centering 2/1 & \centering 0.679 & \centering 0.844 & \centering 0.68/0.70 & \centering 0.96/0.89 & \centering 0.92/0.80 \tabularnewline
\hline
\end{tabular}
\end{table}

\begin{table}[htp]
\footnotesize
\setlength{\tabcolsep}{3.5pt}
\caption{Real minibatch optimal transport (\(B=64\)) against the entropic-transport partner CELLOT(\(\epsilon^*\)) matched to the loss-based estimate of \(S\): \(\epsilon^*/s^2\); mass of the plan on the diagonal; seeds of the partner and of MBR; ratio \(\rho_N\) of the trained median Euler errors at \(N=4\) and 16 (geometric means over training seeds, 2000 starts); exact-field median error of CELLOT over that of DIAG(\(\kappa^*\)) at the same nominal \(S\) (\(N=4\), 16); trained over exact median error of the partner (\(N=4\), 16); and the first-order prediction \(g_{\rm MBR}/g_{\rm CELLOT}\) over the observed ratio (\(N=4\), 16). The last three columns use the first 800 starts.}
\label{tab:15}
\begin{tabular}{p{\dimexpr 0.053\textwidth-0.954\tabcolsep\relax}p{\dimexpr 0.125\textwidth-2.250\tabcolsep\relax}p{\dimexpr 0.074\textwidth-1.332\tabcolsep\relax}p{\dimexpr 0.089\textwidth-1.602\tabcolsep\relax}p{\dimexpr 0.089\textwidth-1.602\tabcolsep\relax}p{\dimexpr 0.089\textwidth-1.602\tabcolsep\relax}p{\dimexpr 0.160\textwidth-2.880\tabcolsep\relax}p{\dimexpr 0.160\textwidth-2.880\tabcolsep\relax}p{\dimexpr 0.161\textwidth-2.898\tabcolsep\relax}}
\hline
\raggedright tgt & \centering \(\epsilon^*/s^2\) & \centering diag & \centering seeds & \centering \(\rho_4\) & \centering \(\rho_{16}\) & \centering ex. C/D & \centering tr./ex. & \centering g/obs. \tabularnewline
\hline
\raggedright 24 & \centering 0.661 & \centering 0.44 & \centering 3/3 & \centering 1.035 & \centering 0.983 & \centering 1.01/1.00 & \centering 1.02/1.02 & \centering 0.94/0.98 \tabularnewline
\raggedright 31 & \centering 0.956 & \centering 0.25 & \centering 3/3 & \centering 0.661 & \centering 0.641 & \centering 1.07/1.13 & \centering 1.01/1.00 & \centering 0.98/0.99 \tabularnewline
\raggedright 28 & \centering 1.558 & \centering 0.23 & \centering 3/3 & \centering 0.704 & \centering 0.664 & \centering 1.06/1.07 & \centering 0.99/1.00 & \centering 0.92/0.98 \tabularnewline
\raggedright 21 & \centering 1.464 & \centering 0.24 & \centering 3/3 & \centering 0.903 & \centering 0.874 & \centering 0.93/0.98 & \centering 1.00/1.00 & \centering 0.95/0.97 \tabularnewline
\raggedright 22 & \centering 1.151 & \centering 0.28 & \centering 2/1 & \centering 0.690 & \centering 0.707 & \centering 1.17/1.22 & \centering 1.05/1.02 & \centering 0.97/0.97 \tabularnewline
\raggedright 29 & \centering 1.868 & \centering 0.16 & \centering 2/1 & \centering 0.228 & \centering 0.200 & \centering 1.25/1.21 & \centering 1.02/1.02 & \centering 0.86/0.99 \tabularnewline
\raggedright 26 & \centering 0.446 & \centering 0.67 & \centering 2/1 & \centering 0.942 & \centering 0.930 & \centering 1.05/1.02 & \centering 1.01/1.03 & \centering 0.92/0.91 \tabularnewline
\raggedright 23 & \centering 1.532 & \centering 0.15 & \centering 2/1 & \centering 0.871 & \centering 0.933 & \centering 0.76/0.86 & \centering 1.03/1.02 & \centering 0.81/0.83 \tabularnewline
\hline
\end{tabular}
\end{table}

\begin{table}[htp]
\footnotesize
\setlength{\tabcolsep}{3.5pt}
\caption{Real minibatch optimal transport (\(B=64\)) in the predictor comparison of trained fields, \(N=16\): \(S\) (loss route) and median \(|g|\) of the trained field relative to the trained IND of the same target, the ratio of the median Euler errors \(D_{16}\), the held-out residual (log) of the fits in \(S\) and in \(g\) at the MBR row, and the per-target ratio of the RMS residuals over the three rows MIX, MB, MBR.}
\label{tab:13}
\begin{tabular}{p{\dimexpr 0.102\textwidth-1.428\tabcolsep\relax}p{\dimexpr 0.215\textwidth-3.010\tabcolsep\relax}p{\dimexpr 0.208\textwidth-2.912\tabcolsep\relax}p{\dimexpr 0.085\textwidth-1.190\tabcolsep\relax}p{\dimexpr 0.136\textwidth-1.904\tabcolsep\relax}p{\dimexpr 0.136\textwidth-1.904\tabcolsep\relax}p{\dimexpr 0.118\textwidth-1.652\tabcolsep\relax}}
\hline
\raggedright target & \centering \(S_{\rm MBR}/S_{\rm IND}\) & \centering \(g_{\rm MBR}/g_{\rm IND}\) & \centering \(D_{16}\) ratio & \centering residual of \(S\) & \centering residual of \(g\) & \centering residual ratio across the three rows \tabularnewline
\hline
\raggedright 21 & \centering 0.085 & \centering 0.050 & \centering 0.054 & \centering +0.312 & \centering -0.014 & \centering 0.036 \tabularnewline
\raggedright 22 & \centering 0.095 & \centering 0.034 & \centering 0.037 & \centering -0.252 & \centering -0.007 & \centering 0.027 \tabularnewline
\raggedright 23 & \centering 0.213 & \centering 0.044 & \centering 0.059 & \centering -0.714 & \centering +0.226 & \centering 0.310 \tabularnewline
\raggedright 24 & \centering 0.090 & \centering 0.059 & \centering 0.062 & \centering +0.387 & \centering -0.036 & \centering 0.089 \tabularnewline
\raggedright 25 & \centering 0.104 & \centering 0.077 & \centering 0.083 & \centering +0.532 & \centering +0.005 & \centering 0.059 \tabularnewline
\raggedright 26 & \centering 0.076 & \centering 0.056 & \centering 0.062 & \centering +0.628 & \centering +0.016 & \centering 0.050 \tabularnewline
\raggedright 27 & \centering 0.088 & \centering 0.046 & \centering 0.054 & \centering +0.258 & \centering +0.066 & \centering 0.239 \tabularnewline
\raggedright 28 & \centering 0.093 & \centering 0.034 & \centering 0.036 & \centering -0.258 & \centering -0.050 & \centering 0.165 \tabularnewline
\raggedright 29 & \centering 0.103 & \centering 0.014 & \centering 0.015 & \centering -1.337 & \centering -0.088 & \centering 0.067 \tabularnewline
\raggedright 30 & \centering 0.102 & \centering 0.049 & \centering 0.051 & \centering +0.024 & \centering -0.050 & \centering 0.294 \tabularnewline
\raggedright 31 & \centering 0.096 & \centering 0.034 & \centering 0.036 & \centering -0.289 & \centering -0.024 & \centering 0.128 \tabularnewline
\hline
\end{tabular}
\end{table}
\begin{table}[htp]

\scriptsize
\setlength{\tabcolsep}{3.5pt}
\caption{Trained two-dimensional MLPs at 20000 steps over all available seeds (see Table~\ref{tab:6} for individual configurations): mean and range over seeds of the relative velocity error, the relative Jacobian error and the density-weighted RMS curl-to-Jacobian ratio.}
\label{tab:6c}
\begin{tabular}{p{\dimexpr 0.072\textwidth-1.008\tabcolsep\relax}p{\dimexpr 0.109\textwidth-1.526\tabcolsep\relax}p{\dimexpr 0.084\textwidth-1.176\tabcolsep\relax}p{\dimexpr 0.060\textwidth-0.840\tabcolsep\relax}p{\dimexpr 0.225\textwidth-3.150\tabcolsep\relax}p{\dimexpr 0.225\textwidth-3.150\tabcolsep\relax}p{\dimexpr 0.225\textwidth-3.150\tabcolsep\relax}}
\hline
\raggedright source & \centering model & \centering curl penalty & \centering seeds & \centering relative \(v\) error mean (range) & \centering relative \(J\) error mean (range) & \centering curl-to-Jacobian RMS ratio mean (range) \tabularnewline
\hline
\raggedright aniso & \centering free & \centering lam0.0 & \centering 0,1,2 & \centering 0.0196 (0.0187 to 0.0201) & \centering 0.1298 (0.1267 to 0.1347) & \centering 0.4103 (0.4092 to 0.4114) \tabularnewline
\raggedright aniso & \centering grad & \centering lam0.0 & \centering 0,1 & \centering 0.1420 (0.1419 to 0.1421) & \centering 0.4110 (0.4104 to 0.4116) & \centering 0.0000 (0.0000 to 0.0000) \tabularnewline
\raggedright iso & \centering free & \centering lam0.0 & \centering 0,1,2 & \centering 0.0187 (0.0182 to 0.0194) & \centering 0.1229 (0.1222 to 0.1244) & \centering 0.0420 (0.0413 to 0.0429) \tabularnewline
\raggedright iso & \centering free & \centering lam1.0 & \centering 0,1 & \centering 0.0184 (0.0179 to 0.0190) & \centering 0.1270 (0.1239 to 0.1301) & \centering 0.0080 (0.0077 to 0.0083) \tabularnewline
\raggedright iso & \centering grad & \centering lam0.0 & \centering 0,1,2 & \centering 0.0160 (0.0155 to 0.0168) & \centering 0.1100 (0.1058 to 0.1137) & \centering 0.0000 (0.0000 to 0.0000) \tabularnewline
\raggedright iso & \centering gradfloor & \centering lam0.0 & \centering 0 & \centering 0.0048 (0.0048 to 0.0048) & \centering 0.0459 (0.0459 to 0.0459) & \centering 0.0000 (0.0000 to 0.0000) \tabularnewline
\raggedright mix & \centering free & \centering lam0.0 & \centering 0,1,2 & \centering 0.0241 (0.0237 to 0.0244) & \centering 0.1432 (0.1427 to 0.1441) & \centering 0.3754 (0.3722 to 0.3776) \tabularnewline
\raggedright mix & \centering free & \centering lam1.0 & \centering 0,1 & \centering 0.1243 (0.1239 to 0.1248) & \centering 0.3752 (0.3742 to 0.3761) & \centering 0.0275 (0.0269 to 0.0280) \tabularnewline
\raggedright mix & \centering grad & \centering lam0.0 & \centering 0,1,2 & \centering 0.1359 (0.1358 to 0.1360) & \centering 0.3869 (0.3864 to 0.3880) & \centering 0.0000 (0.0000 to 0.0000) \tabularnewline
\raggedright mix & \centering gradfloor & \centering lam0.0 & \centering 0 & \centering 0.1347 (0.1347 to 0.1347) & \centering 0.3760 (0.3760 to 0.3760) & \centering 0.0000 (0.0000 to 0.0000) \tabularnewline
\hline
\end{tabular}
\end{table}

\begin{table}[htp]
\footnotesize
\setlength{\tabcolsep}{3.5pt}
\caption{Trained two-dimensional MLPs (3 hidden layers of 128, 20000 Adam steps, flow-matching loss on fresh samples), density-weighted over \(t=0.1,\dots,0.9\) against the exact closed-form field. The first column gives the source, model, seed and penalty weight. The model and exact RMS ratios divide RMS curl by RMS \(\|J\|_F\).}
\label{tab:6}
\begin{tabular}{p{\dimexpr 0.318\textwidth-4.452\tabcolsep\relax}p{\dimexpr 0.111\textwidth-1.554\tabcolsep\relax}p{\dimexpr 0.111\textwidth-1.554\tabcolsep\relax}p{\dimexpr 0.129\textwidth-1.806\tabcolsep\relax}p{\dimexpr 0.092\textwidth-1.288\tabcolsep\relax}p{\dimexpr 0.147\textwidth-2.058\tabcolsep\relax}p{\dimexpr 0.092\textwidth-1.288\tabcolsep\relax}}
\hline
\raggedright Model configuration & \centering relative \(v\) error & \centering relative \(J\) error & \centering RMS curl/\(\|J\|_F\) & \centering relative curl error & \centering antisym. share of \(J\) error & \centering exact RMS ratio \tabularnewline
\hline
\raggedright Isotropic FREE, 0, \(\lambda=0.0\) & \centering 0.0185 & \centering 0.1222 & \centering 0.0429 & \centering - & \centering 0.061 & \centering 0 \tabularnewline
\raggedright Isotropic FREE, 1, \(\lambda=0.0\) & \centering 0.0182 & \centering 0.1244 & \centering 0.0417 & \centering - & \centering 0.056 & \centering 0 \tabularnewline
\raggedright Isotropic GRAD, 0, \(\lambda=0.0\) & \centering 0.0155 & \centering 0.1058 & \centering 0.0000 & \centering - & \centering 0.000 & \centering 0 \tabularnewline
\raggedright Isotropic GRADFLOOR, 0, \(\lambda=0.0\) & \centering 0.0048 & \centering 0.0459 & \centering 0.0000 & \centering - & \centering 0.000 & \centering 0 \tabularnewline
\raggedright Mixture FREE, 0, \(\lambda=0.0\) & \centering 0.0244 & \centering 0.1441 & \centering 0.3722 & \centering 0.196 & \centering 0.130 & \centering 0.375 \tabularnewline
\raggedright Mixture GRAD, 0, \(\lambda=0.0\) & \centering 0.1358 & \centering 0.3880 & \centering 0.0000 & \centering 1.000 & \centering 0.466 & \centering 0.375 \tabularnewline
\raggedright Mixture GRADFLOOR, 0, \(\lambda=0.0\) & \centering 0.1347 & \centering 0.3760 & \centering 0.0000 & \centering 1.000 & \centering 0.496 & \centering 0.375 \tabularnewline
\raggedright Anisotropic FREE, 0, \(\lambda=0.0\) & \centering 0.0187 & \centering 0.1280 & \centering 0.4114 & \centering 0.157 & \centering 0.127 & \centering 0.409 \tabularnewline
\raggedright Anisotropic GRAD, 0, \(\lambda=0.0\) & \centering 0.1421 & \centering 0.4116 & \centering 0.0000 & \centering 1.000 & \centering 0.494 & \centering 0.409 \tabularnewline
\raggedright Isotropic FREE, 0, \(\lambda=1.0\) & \centering 0.0179 & \centering 0.1239 & \centering 0.0077 & \centering - & \centering 0.002 & \centering 0 \tabularnewline
\raggedright Mixture FREE, 0, \(\lambda=1.0\) & \centering 0.1248 & \centering 0.3761 & \centering 0.0269 & \centering 0.954 & \centering 0.451 & \centering 0.375 \tabularnewline
\raggedright Mixture FREE, 1, \(\lambda=0.0\) & \centering 0.0237 & \centering 0.1428 & \centering 0.3776 & \centering 0.199 & \centering 0.137 & \centering 0.375 \tabularnewline
\hline
\end{tabular}
\end{table}
\begin{table}[htp]
\small
\setlength{\tabcolsep}{3.5pt}
\caption{Vorticity during training with an isotropic source. Ratios use density-weighted RMS norms and \(\Omega=(J-J^T)/2\).}
\label{tab:6b}
\begin{tabular}{p{\dimexpr 0.156\textwidth-1.560\tabcolsep\relax}p{\dimexpr 0.187\textwidth-1.870\tabcolsep\relax}p{\dimexpr 0.187\textwidth-1.870\tabcolsep\relax}p{\dimexpr 0.218\textwidth-2.180\tabcolsep\relax}p{\dimexpr 0.252\textwidth-2.520\tabcolsep\relax}}
\hline
\raggedright step & \centering relative \(v\) error & \centering relative \(J\) error & \centering RMS curl/\(\|J\|_F\) & \centering RMS ratio \(J/\Omega\) \tabularnewline
\hline
\raggedright 500 & \centering 0.1526 & \centering 0.4898 & \centering 0.1161 & \centering 12.2 \tabularnewline
\raggedright 2000 & \centering 0.0806 & \centering 0.2815 & \centering 0.0968 & \centering 14.6 \tabularnewline
\raggedright 6000 & \centering 0.0539 & \centering 0.1852 & \centering 0.0753 & \centering 18.8 \tabularnewline
\raggedright 12000 & \centering 0.0380 & \centering 0.1492 & \centering 0.0544 & \centering 26.0 \tabularnewline
\raggedright 20000 & \centering 0.0185 & \centering 0.1222 & \centering 0.0429 & \centering 33.0 \tabularnewline
\hline
\end{tabular}
\end{table}

\begin{thebibliography}{99}
\small
\bibitem{r1} Lipman, Chen, Ben-Hamu, Nickel, Le, ``Flow matching for generative modeling'', ICLR (2023), arXiv:2210.02747.
\bibitem{r2} Liu, Gong, Liu, ``Flow straight and fast: learning to generate and transfer data with rectified flow'', ICLR (2023), arXiv:2209.03003.
\bibitem{r3} Tong, Fatras, Malkin, Huguet, Zhang, Rector-Brooks, Wolf, Bengio, ``Improving and generalizing flow-based generative models with minibatch optimal transport'', TMLR (2024), arXiv:2302.00482.
\bibitem{r4} Pooladian, Ben-Hamu, Domingo-Enrich, Amos, Lipman, Chen, ``Multisample flow matching: straightening flows with minibatch couplings'', ICML (2023), pp.~28100--28127.
\bibitem{r5} Lee, Kim, Ye, ``Minimizing trajectory curvature of ODE-based generative models'', ICML (2023), pp.~18957--18973.
\bibitem{r6} Tao, Choi, ``On the role of strain and vorticity in numerical integration error for flow matching'', arXiv:2605.06680 (2026).
\bibitem{r7} Xing, Wang, Wang, \emph{The divergence is the uncertainty: a closed-form identity for flow matching}, arXiv:2605.00941v3 (2026), Eq.~2 and Corollary 3.
\bibitem{r11} Yin, ``Spatial transport of integration error in generative ODEs'', arXiv:2607.16361 (2026).
\bibitem{r12} Chao, Sun, Cheng, Lee, ``On investigating the conservative property of score-based generative models'', arXiv:2209.12753 (2022).
\bibitem{r13} Praneeth, Shanker, Dubey, Seelamantula, \emph{Velocity adaptation for flow-matching models}, ICML FoGen Workshop (2026), Sec.~4, \url{https://openreview.net/forum?id=tdIwEN0fCi}.
\bibitem{r14} Binney, Tremaine, \emph{Galactic Dynamics}, 2nd ed., Princeton University Press (2008), Sec.~4.8.
\bibitem{r15} Khan, ``Isokinetic flow matching for pathwise straightening of generative flows'', arXiv:2604.04491 (2026), Appendix A.5, Eqs. 25 to 35 (p.~13).
\bibitem{r16} Lim, ``On the hidden biases of flow matching samplers'', arXiv:2512.16768v3 (2026).
\bibitem{r17} Obzherin, Khrulkov, ``Density-level backward error for low-NFE generative flows'', arXiv:2610.29475 (2026), Theorems 1--3.
\bibitem{r18} Obzherin, Khrulkov, ``Finite-step geometry of marginal-preserving velocity fields in generative flows'', arXiv:2610.37936v1 (2026).
\bibitem{r19} Cuturi, ``Sinkhorn distances: lightspeed computation of optimal transportation distances'', arXiv:1306.0895 (2013), Sec. 4.
\bibitem{r20} Tsimpos, Marzouk, \emph{An Eulerian perspective on straight-line sampling}, arXiv:2510.11657v2 (2025), Lemma 3 and Theorem 6.
\bibitem{r21} Sun, Samaddar, Madireddy, \emph{Composing flow-matching energies with known physics: generation, OOD detection, and inversion on PDE fields}, arXiv:2608.18004 (2026), Eq.~7.
\bibitem{r22} Peng, Ma, \emph{Classifier-free guidance in flow matching: non-autonomous potentials, overshoot, and posterior-mean control}, arXiv:2609.24287 (2026), Proposition 3.1 and Appendix A.
\bibitem{r23} Zhang, Huang, Yu, Li, \emph{Potential matching optimal transport: continuous normalizing flows for exact \(p\)-Wasserstein dynamics}, arXiv:2608.05666 (2026).
\bibitem{r24} Hairer, N{\o}rsett, Wanner, \emph{Solving Ordinary Differential Equations I: Nonstiff Problems}, 2nd ed., Springer (1993), Sec.~II.8.
\bibitem{pham} Pham, Roy, Rinaldo, Sarkar, \emph{Setting things straight: on the straightness and non-straightness of two-step rectified flow}, arXiv:2610.36090v2 (2026).
\bibitem{claustre} Claustre, Negrel, Boyer, Nadjahi, Vanden-Eijnden, \emph{Gaussian flow-matching schedules: implications for sampling and training}, arXiv:2609.25839 (2026).
\bibitem{boite} Bo\"it\'e, Delon, Nadjahi, \emph{Expected batch optimal transport plans and consequences for flow matching}, arXiv:2605.12174v2 (2026).
\bibitem{chen} Chen, Levine, \emph{Scaling online optimal transport minibatches for flow matching}, arXiv:2610.18838v1 (2026).
\bibitem{gribonval} Gribonval, Nikolova, \emph{On Bayesian estimation and proximity operators}, Appl. Comput. Harmon. Anal. \textbf{50}, 49--72 (2021), \href{https://doi.org/10.1016/j.acha.2019.07.002}{doi:10.1016/j.acha.2019.07.002}.
\bibitem{gao} Gao, Huang, Jiao, \emph{Gaussian interpolation flows}, J. Mach. Learn. Res. \textbf{25}(253), 1--52 (2024).
\bibitem{dytso} Dytso, Poor, Shamai, \emph{Conditional mean estimation in Gaussian noise: a meta derivative identity with applications}, IEEE Trans. Inf. Theory \textbf{69}, 1883--1898 (2023).
\bibitem{albergo} Albergo, Boffi, Vanden-Eijnden, \emph{Stochastic interpolants: a unifying framework for flows and diffusions}, J. Mach. Learn. Res. \textbf{26}(209), 1--80 (2025).
\bibitem{bansal} Bansal, Roy, Rinaldo, Sarkar, \emph{On the convergence and straightness of rectified flow}, arXiv:2410.14949v8 (2026).
\bibitem{hertrich} Hertrich, Chambolle, Delon, \emph{On the relation between rectified flows and optimal transport}, NeurIPS (2025).
\bibitem{wan} Wan, Wang, Mishne, Wang, \emph{Elucidating flow matching ODE dynamics via data geometry and denoisers}, ICML (2025).
\end{thebibliography}
\end{document}